# Prospects of ratio and differential (δ) ratio based measurement-models: a case study for IRMS evaluation

*B. P. Datta* (email: bibek@vecc.gov.in)
Radiochemistry Laboratory, Variable Energy Cyclotron Centre, 1/AF Bidhan Nagar, Kolkata 700 064, India

**ABSTRACT**

The suitability of a mathematical-model "$\boldsymbol{Y = f(\{X_i\})}$" in serving a purpose whatsoever (should be preset by the function "$\boldsymbol{f}$" specific input-to-output variation-rates, i.e.) can be judged beforehand. We thus evaluate here the two apparently similar models "$\boldsymbol{Y_A = f_A({}^{S}R_i, {}^{W}R_i) = ({}^{S}R_i/{}^{W}R_i)}$" and "$\boldsymbol{Y_\delta = f_\delta({}^{S}R_i, {}^{W}R_i) = ([{}^{S}R_i, {}^{W}R_i] - 1) = (Y_A - 1)}$", with $\boldsymbol{{}^{S}R_i}$ and $\boldsymbol{{}^{W}R_i}$ representing certain measurable-variables (e.g. the sample $\boldsymbol{S}$ and the working-lab-reference $\boldsymbol{W}$ specific $i^{th}$-isotopic-abundance-ratios, respectively, for a case as the isotope ratio mass spectrometry "IRMS"). The idea is to ascertain whether "$\boldsymbol{f_\delta}$" should represent a better model than "$\boldsymbol{f_A}$", specifically, for the well-known IRMS evaluation.

The study clarifies that "$\boldsymbol{f_A}$" and "$\boldsymbol{f_\delta}$" should really represent different model-families. For example, the possible variation, $\boldsymbol{\varepsilon_A}$, of an *absolute* estimate as the "$\boldsymbol{y_A}$" (and/ or the risk of running a machine on the basis of the measurement-model "$\boldsymbol{f_A}$") should be dictated by the possible $\boldsymbol{R_i}$-measurement-variations ($\boldsymbol{u_S}$ and $\boldsymbol{u_W}$) only: $\boldsymbol{\varepsilon_A = (\sum_i u_i) = (u_S + u_W)}$; i.e., at worst: $\boldsymbol{\varepsilon_A = 2u_i}$. However, the variation, $\boldsymbol{\varepsilon_\delta}$, of the corresponding *differential* (i.e. $\boldsymbol{{}^{S/W}\delta_i \equiv Y_\delta}$) estimate "$\boldsymbol{y_\delta}$" should largely be decided by "$\boldsymbol{{}^{S}R_i}$ and $\boldsymbol{{}^{W}R_i}$" values: $\boldsymbol{\varepsilon_\delta = \sum_i(|m_i| \times u_i) = 2(|m_i| \times u_i) = (|m_i| \times \varepsilon_A)}$; with: $\boldsymbol{m_i = ({}^{S}R_i/[{}^{S}R_i - {}^{W}R_i])}$.

Thus, any IRMS measurement (i.e. for which "$\boldsymbol{|{}^{S}R_i - {}^{W}R_i| \to 0}$" is a requirement) should signify that "$\boldsymbol{|m_i| \to \infty}$". Clearly, "$\boldsymbol{y_\delta}$" should be less accurate than "$\boldsymbol{y_A}$", and/ or even turn out to be highly erroneous ($\boldsymbol{\varepsilon_\delta \to \infty}$). Nevertheless, the evaluation as the absolute-ratio "$\boldsymbol{y_A}$", and hence as the sample isotopic ratio "${}^{S}r_i$", is shown to be equivalent to our previously reported finding that the conversion of a **δ**-estimate (here, $y_\delta$) into "${}^{S}r_i$" should help to improve the achievable output-accuracy and -comparability.

## 1. INTRODUCTION

Evaluation should in general mean knowing any unknown (e.g. value of a variable, or relationship between variables, or **…**) whatever. However, any conceivable result should be based on some defined standard(s) and/ or knowledge, i.e. refer to a relative fact only. For example, the understanding of a result ("$x$" Kg) of even simply weighing an unknown amount ("$X$" Kg of a solid-material) should need the 'a priori' knowledge of weighing-unit "Kg". Further, the accountability of any estimate-of-*indirect*-measurement ($y$) should require the additional knowledge of relationship (viz.**:** $Y = f(\{X_i\})$) of desired variable ($Y$) with corresponding measurable variables ($X_i$, $i$ = 1, 2 **…**). However, the relationship (function "$f$") may also happen to be unknown, and/ or be a *proposed* one. The latter, i.e. a *possible* relationship between any known (measurable) and unknown (desired) variables, is generally referred to as a "mathematical-model". Again, for a given purpose, there might be several proposals. Thus, e.g. the *comparison* between two different quantities, $X_J$ and $X_K$, might be carried out in terms of their *absolute* ratio ($Y_A$) or *differential* ratio[1] ($Y_\delta \equiv \delta$):

$$Y_A = f_A(X_J, X_K) = (X_J/X_K) \quad (1)$$

$$Y_\delta = f_\delta(X_J, X_K) = ([X_J/X_K] - 1) = (Y_A - 1) \quad (2)$$

However, for any specified purpose whatever, how should really a proper model (or, if applicable, an appropriate *derivable* formula from many a possible one) be chosen?

This work considers the evaluation system to be the well-known isotope ratio mass spectrometry (IRMS),[1,2] and discusses a simple means for resolving the issue.

## 2. TERMINOLOGIES AND PRINCIPLES

It is important pointing out that, irrespective of purpose (here, IRMS evaluation), any proposed model ("$\boldsymbol{f_A}$" or "$\boldsymbol{f_\delta}$") should represent a specific method for bringing out a net *systematic*[3] change in the corresponding independent variables (here: $\boldsymbol{X_J}$ and $\boldsymbol{X_K}$) and/ or estimates ("$\boldsymbol{x_J} \pm \boldsymbol{u_J}$" and "$\boldsymbol{x_K} \pm \boldsymbol{u_K}$"), and hence for dictating (the *nature* of) the *output* to be expected. Therefore, 'a priori' study of the possible properties[3] of a proposed input-output relationship (e.g. rates of variations of the modeled variable "$\boldsymbol{Y_A}$" or "$\boldsymbol{Y_\delta}$" as a function of the measurable variables "$\boldsymbol{X_J}$ and $\boldsymbol{X_K}$") should be a means for proper modeling.

### 2.1 Error and Uncertainty

As well-known[4], the process "*measurement*" is subject to error. That is, even a directly measurable estimate "$\boldsymbol{x_i}$" could be different from the corresponding *unknown* true-value "$\boldsymbol{X_i}$"; e.g.: $\boldsymbol{X_i} = (\boldsymbol{x_i} + \Delta_i)$, with $\Delta_i$ representing the error (if any) in "$\boldsymbol{x_i}$". Thus, assessment of error should be an integral part of any evaluation. However, ascertainment of a true error as "$\Delta_i$" is inconceivable, and any result is generally reported as: $\boldsymbol{X_i} = (\boldsymbol{x_i} \pm \boldsymbol{u_i})$. Clearly, "$\boldsymbol{u_i}$" should represent the *possible* and, therefore the ***maximum***, *value* of the error "$\Delta_i$" (i.e.[3]: $\boldsymbol{u_i} = {}^{\mathbf{Max}}|\Delta_i|$); and be ensured 'a priori' (by the aid of relevant standards, i.e. while developing a chosen measurement technique) to be, at least acceptably, small.

Similarly, any indirect-measurement *model* "$\boldsymbol{Y_d} = \boldsymbol{f_d}(\{\boldsymbol{X_i}\}_{\boldsymbol{i=1}}^{\boldsymbol{N}})$" should in terms of corresponding *independent* (measurable input) and *dependent* (desired output) estimates be represented as: $(\boldsymbol{y_d} + \boldsymbol{Đ_d}) = \boldsymbol{f_d}(\{\boldsymbol{x_i} + \Delta_i\}_{\boldsymbol{i=1}}^{\boldsymbol{N}})$; and/ or: $(\boldsymbol{y_d} \pm \boldsymbol{\varepsilon_d}) = \boldsymbol{f_d}(\{\boldsymbol{x_i} \pm \boldsymbol{u_i}\}_{\boldsymbol{i=1}}^{\boldsymbol{N}})$; where "$\text{Đ}_d$" should stand for true *modeling* (output) error, and $\varepsilon_d$ for the corresponding *maximum possible value* (**MPV**); i.e.: $\varepsilon_d = {}^{\text{Max}}|\text{Đ}_d|$.

However, it could also be mentioned that: **(i)** *method-development* may not always help to even identify small sources of systematic errors; i.e. even a *small* measurement-error ($\Delta_i$) may not be purely random by origin; and: **(ii)** any modeling error "$\text{Đ}_d$" should, in nature, be purely systematic[3]: $\text{Đ}_d = f_d(\{\Delta_i\}_{i=1}^{N})$. Moreover, "$\text{Đ}_d$" should ***never*** vary for whether "$\Delta_i$" should be purely random or purely systematic or both by the origin. Thus, unlike the ref. [4], we refer to the "**MPV** of any (direct/ indirect measurement) error" as[3] the (corresponding) *accuracy* or inaccuracy or ***uncertainty***.

Further, only relative-error should be the measure of an error. Thus, by the error "$\Delta_i$", we do mean that: $\Delta_i = \frac{\Delta X_i}{X_i} = \frac{x_i - X_i}{X_i}$. Similarly, we define (modeling/ output error): $\text{Đ}_d = \frac{dY_d}{Y_d} = \frac{y_d - Y_d}{Y_d}$. Therefore, by the uncertainty "$u_i$" or "$\varepsilon_d$", we refer to the corresponding "*relative* **MPV**" only.

## 2.2 Evaluation of models

For any model "$Y_d = f_d(\{X_i\}_{i=1}^{N})$", and hence for any modeled estimate: $(y_d + \text{Đ}_d) = f_d(\{x_i + \Delta_i\}_{i=1}^{N})$; and/ or: $(y_d \pm \varepsilon_d) = f_d(\{x_i \pm u_i\}_{i=1}^{N})$; the modeling-error ($\text{Đ}_d$) could be shown to be decided as[3,5]:

$$\text{Đ}_d = \sum_{i=1}^{N} (M_i^d \times \Delta_i) \tag{3}$$

And, the modeling-uncertainty ($\varepsilon_d$) can, really a priori, be ascertained as[3,6]:

$$\varepsilon_d = \sum_{i=1}^{N} (|M_i^d| \times u_i) = ([UF]_d \times {}^{G}u) \tag{4}$$

where $M_i^d$ is a theoretical constant, representing the model ("$f_d$") specific relative rate of variation of the modeled variable $Y_d$ as a function of the measurable variable $X_i$:

$$M_i^d = \left(\frac{\partial Y_d}{\partial X_i}\right)\left(\frac{X_i}{Y_d}\right) = \left(\frac{\partial Y_d/Y_d}{\partial X_i/X_i}\right), \quad i = 1, 2 \ldots N \tag{5}$$

And, ${}^{G}\boldsymbol{u}$ stands for the measurement accuracy to be achieved **(**i.e. which is generally preset before developing a required experimental-methodology and/ or ascertaining the possible $\boldsymbol{X_i}$ specific "$\boldsymbol{u_i}$"**)**; so that**:** $\boldsymbol{u_i}$ = ${}^{G}\boldsymbol{u}$ (with**:** $\boldsymbol{i}$ = 1, 2 **…** $N$); and:

$$[UF]_d = (\varepsilon_d/{}^{G}u) = \sum_{i=1}^{N}\left|M_i^d\right| \tag{6}$$

Thus "$[\boldsymbol{UF}]_{\boldsymbol{d}}$", which should be referred to[6] as the *uncertainty-factor* of modeling (determining $\boldsymbol{Y_d}$), really represents the *collective* rate-of variations of $\boldsymbol{Y_d}$ as a function of *all* different input/ measurable variables ($\boldsymbol{X_i}$, with**:** $\boldsymbol{i}$ = 1, 2 **…** $\boldsymbol{N}$). Therefore, the smaller should be the value of $[\boldsymbol{UF}]_{\boldsymbol{d}}$ (viz.**:** $[\boldsymbol{UF}]_{\boldsymbol{d}} < \mathbf{1}$, rather**:** $[\boldsymbol{UF}]_{\boldsymbol{d}} \ll \mathbf{1}$) the better be the (evaluation method represented by the proposed) model "$\boldsymbol{f_d}$".

### 2.2.1 *Clues for modeling: nature of parameters to be looked for*

We may here go backward by examining certain formulae, e.g. Eqs 1 and 2, for their behavior.

"Eq. 1**:** $\boldsymbol{Y_A = (X_J/X_K)}$" is, however, already shown elsewhere[3] to be characterized by the following parameters (cf. Eq. 5)**:** $\boldsymbol{M_J^A} = \mathbf{1}$, and**:** $\boldsymbol{M_K^A} = -\mathbf{1}$; and thus (cf. Eq. 3)**:**

$$Đ_A = \sum_{i=1}^{2} (M_i^A \times \Delta_i) = (\Delta_J - \Delta_K) \tag{3a}$$

And (cf. Eq. 4)

$$\varepsilon_A = \sum_{i=1}^{2} (\left|M_i^A\right| \times u_i) = (u_J + u_K) \tag{4a}$$

Further (cf. Eq. 6):

$$[UF]_A = \sum_{i=1}^{2} (\left|M_i^A\right|) = (\left|M_J^A\right| + |M_K^A|) = 2 \tag{6a}$$

Therefore (for**:** $\boldsymbol{u_J} = \boldsymbol{u_K} = {}^{G}\boldsymbol{u}$; cf. Eq. 4 and/ or 4a):

$$\varepsilon_A = ([UF]_A \times {}^{G}u) = (2 \times {}^{G}u) \tag{4a$'$}$$

Thus, what is signified is that even any elementary mathematical process should *not* be presumed to be (characterized by "$[\boldsymbol{UF}]_{\boldsymbol{d}}$ **= 1**", and hence) *non-biasing*. For example, any *estimated* unknown ratio "$\boldsymbol{y_A}$" (and hence any $\boldsymbol{y_A}$ based *insight*) could be twice as *wrong* as a corresponding *monitored* estimate "$\boldsymbol{x_J}$ or $\boldsymbol{x_K}$".

Yet, it is important pointing out that the "$\boldsymbol{Y_A = (X_J/X_K)}$" specific parameters (viz. the *individual* rates-of-variations "$\boldsymbol{M_J^A}$ and, $\boldsymbol{M_K^A}$", and thus the *collective* rate-of-variation "$[\boldsymbol{UF}]_{\boldsymbol{A}}$" and/ or the achievable *output-accuracy* "$\varepsilon_{\boldsymbol{A}}$") should all be ***independent*** of the measureable variables "$\boldsymbol{X_J}$ and $\boldsymbol{X_K}$". Again, any analytical method should be valuable provided the corresponding desired result can 'a priori', i.e. *irrespective* of *what* and *how much* to be measured, be assured to be accurate. Thus, "$\boldsymbol{Y_A = (X_J/X_K)}$" may stand out to represent a good evaluation model. Further, it could be shown below that "$\boldsymbol{Y_A = (X_J/X_K)}$" helps minimize the effect of possible, i.e. even *undetectable* and/ or *uncorrectable*, systematic errors (say, $^{\mathbf{(SYS)}}\boldsymbol{\Delta_J}$ and $^{\mathbf{(SYS)}}\boldsymbol{\Delta_K}$) of measuring "$\boldsymbol{X_J}$ and $\boldsymbol{X_K}$" (respectively) on the modeled estimate "$\boldsymbol{y_A}$", and hence, to improve the desired output accuracy ($\varepsilon_{\boldsymbol{A}}$) as:

$$\text{“}([^{(SYS)}u_J + {}^{(RAN)}u_J] + [^{(SYS)}u_K + {}^{(RAN)}u_K]) \equiv (u_J + u_K)\text{”} > \varepsilon_A \geq \text{“}(^{(RAN)}u_J + {}^{(RAN)}u_K)\text{”}$$

We now refer to the Eq. 2 **(**"$\boldsymbol{Y_\delta = [(X_J/X_K) - 1]} = [\boldsymbol{Y_A - 1}]$", i.e. which signifies that the subtraction of an estimate as "$\boldsymbol{y_A}$" from merely a constant "**1**" should yield the differential estimate "$\boldsymbol{y_\delta}$"**)**, and enquire**:** should "$\boldsymbol{y_\delta}$" be equally as accurate as "$\boldsymbol{y_A}$"?

However, "$\boldsymbol{Y_\delta = [[(X_J/X_K] - 1)}$" could be seen to be characterized by different ***kinds*** of parameters (cf. Eq. 5) as**:** $\boldsymbol{M_J^\delta = \frac{X_J}{X_J - X_K}}$, and**:** $\boldsymbol{M_K^\delta = \frac{-X_J}{X_J - X_K}}$; i.e. to imply that (cf. Eq. 3)**:**

$$Đ_\delta = \sum_{i=1}^{2} (M_i^\delta \times \Delta_i) = \left(\left[\frac{X_J}{X_J - X_K} \times \Delta_J\right] - \left[\frac{X_J}{X_J - X_K} \times \Delta_K\right]\right) \qquad (3b)$$

And (cf. Eq. 4 and also Eq. 4a)

$$\varepsilon_\delta = \sum_{i=1}^{2} (|M_i^\delta| \times u_i) = \left(\left[\left|\frac{x_J}{x_J - x_K}\right| \times u_J\right] + \left[\left|\frac{-x_J}{x_J - x_K}\right| \times u_K\right]\right)$$

$$= \left(\left|\frac{x_J}{x_J - x_K}\right| \times [\mathrm{u_J} + \mathrm{u_K}]\right) = \left(\left|\frac{x_J}{x_J - x_K}\right| \times \boldsymbol{\varepsilon_A}\right) \qquad (4b)$$

Moreover (cf. Eq. 6):

$$[UF]_\delta = \left(\left|\frac{x_J}{x_J - x_K}\right| + \left|\frac{-x_J}{x_J - x_K}\right|\right) = \left(\left|\frac{x_J}{x_J - x_K}\right| \times 2\right) = \left(\left|\frac{x_J}{x_J - x_K}\right| \times \boldsymbol{[UF]_A}\right) \qquad (6b)$$

Thus, even for "$\boldsymbol{u_J} = \boldsymbol{u_K} = {}^{G}\boldsymbol{u}$" (cf. Eq. 4b and also Eq. 4a$^{/}$):

$$\boldsymbol{\varepsilon_\delta} = \left(\left|\frac{x_J}{x_J - x_K}\right| \times 2 \times {}^{G}u\right) = \left(\left|\frac{x_J}{x_J - x_K}\right| \times \boldsymbol{\varepsilon_A}\right) \qquad (4b^{/})$$

Furthermore, the behavior of Eq. 2 might be studied as simply "$\boldsymbol{Y_\delta = (Y_A - 1)}$", i.e. "$\boldsymbol{y_\delta}$" can be evaluated as the **2nd** *stage-estimate*[6]:

$$(y_\delta \pm \varepsilon_\delta) = ([y_A \pm \varepsilon_A] - 1) \qquad (2^{/})$$

where, clearly, $\boldsymbol{y_A}$ should represent the **1st** *stage* estimate (cf. Eq. 1 and also Eq. 4a or 4a$^{/}$):

$$(y_A \pm \varepsilon_A) = \left( [x_J \pm u_J] \big/ [x_K \pm u_K] \right) \qquad (1^{/})$$

However, the change of path should, *on its own*, never make the result to be different. For example, the rate of variation of the ***differential***-*ratio* ($\boldsymbol{Y_\delta}$) as a function of the ***absolute***-*ratio* ($\boldsymbol{Y_A}$), i.e. the **2nd** *stage* output-variation, could be shown to be decided (cf. Eq. 5) as**:** $\boldsymbol{M_A^\delta} = \boldsymbol{\frac{Y_A}{Y_A - 1}} = \boldsymbol{\frac{X_J}{X_J - X_K}}$. Therefore, the **2nd** *stage*[6] output-uncertainty should be governed as **(**cf. Eq. 4 and also Eqs. (4a$^{/}$, 6a, 6b and so)**)**:

$$\varepsilon_\delta = (|M_A^\delta| \times \boldsymbol{\varepsilon_A}) = \left(\left|\frac{x_J}{x_J - x_K}\right| \times \boldsymbol{\varepsilon_A}\right) = \left(\left|\frac{x_J}{x_J - x_K}\right| \times \boldsymbol{[UF]_A} \times {}^{G}u\right)$$

$$= \left(\left|\frac{x_J}{x_J - x_K}\right| \times [2 \times {}^{G}u]\right) = ([UF]_\delta \times {}^{G}u) \qquad (4b^{//})$$

What may however be emphasized is that, on the one, the achievable accuracy ($\varepsilon_A$) of determining any ***absolute*** ratio ($Y_A$) should be governed by the achievable measurement-accuracies ($u_J$ and $u_K$) only, i.e. the measureable variables ($X_J$ and $X_K$) can in no way help in presetting $\varepsilon_A$. On the other, the achievable accuracy ($\varepsilon_\delta$) of determining a **δ-*ratio*** as $Y_\delta$ should largely be *fixed* by "$X_J$ and $X_K$" themselves, rather by the difference "$|X_J - X_K|$". For example, "($|X_J - X_K|$) → **0**" should mean that "$\varepsilon_\delta \rightarrow \infty$", i.e. (for a case, corresponding to which "$y_A$" should rather be *accurate*) the **δ**-estimate "$y_\delta$" can turn out to be highly *erroneous*.

Thus, if "$Y_A = (X_J/X_K)$" and all other such models with "$|M_i^d| = 1$, ($i = 1, 2 \ldots$)" should constitute the family[3] no. **F.1**; then "$Y_\delta = ([X_J/X_K] - 1)$" should belong to another family (no. **F.2**). However, the implication of the family consideration is that, only for any possible "**F.1**" member, the output-accuracy and/ or modeling-performance should be independent of measureable variable(s) "$X_i(s)$". On the contrary, the success of "**F.2**" modeling should be dictated by "$X_i(s)$"; e.g. the model "$Y_\delta = ([X_J/X_K] - 1)$" might lead to disaster in cases as "($|X_J - X_K|$) → **0**" but help exert strong control (i.e. yield even unexpectedly good results) in cases where "($|X_J - X_K|$) → ∞".

## 3 IRMS AND THE RATIO MODELS

The isotopic analysis of any lighter element, rather the study of possible variation in a corresponding source ($S$) specific isotopic-abundance-ratio (${}^S R_i$), is since long[1] proposed and continued to be carried out as a *differential ratio* (**δ**), viz.**:** (${}^{S/W}\delta_i \equiv Y_\delta$) = ($[{}^S R_i/{}^W R_i] - 1$); with $W$ representing a similar isotopic-source as $S$ so that "$|{}^S R_i - {}^W R_i| \rightarrow 0$". The technique of **δ**-measurement is, although lately supplemented by the laser mass spectrometry,[7-9] well-known as the IRMS. Usually, $W$ stands for a relevant lab-available material, and is thus known as the working-lab-reference. However, as different lab-specific ${}^{S/W}\boldsymbol{\delta}_i$-estimates cannot be inter-compared, any species-specific result is reported with reference to a corresponding recommended[10,11] (and hence, say, desired) standard $\boldsymbol{D}$. Thus, while "$\boldsymbol{Y}_\delta$" stands for the (*measurable* and/ or) *IRMS* variable, the *desired* variable is defined to be "$({}^{S/D}\boldsymbol{\delta}_i \equiv \boldsymbol{Z}_\delta) = ([{}^{S}\boldsymbol{R}_i/{}^{D}\boldsymbol{R}_i] - 1)$", with (in principle): ${}^{D}\boldsymbol{R}_i \approx {}^{S}\boldsymbol{R}_i \approx {}^{W}\boldsymbol{R}_i$.

However, why should at all the sample ($\boldsymbol{S}$) measurement be accomplished *by comparison* with another similar isotopic material ($\boldsymbol{W}$ or, $\boldsymbol{D}$), and that too as a **δ**-*variable*?

### 3.1 Measurement by comparison: why?

Any element, specifically a lighter one, should be subject to isotopic fractionation; i.e. isotopic abundances might vary as a function of geo/ bio/ environmental changes. Thus, by the isotopic analysis of an element, it should firstly mean the measurement of relevant *isotopic-abundance-ratios* ($\boldsymbol{R}_i$, $\boldsymbol{i}$ = 1, 2 **…**). Secondly, it is a requirement that "$\boldsymbol{R}_i$," should remain invariant as a function of experimental conditions. However, in case of a lighter element, "$\boldsymbol{R}_i$" (or even simply its estimate "$\boldsymbol{r}_i$") might significantly vary as a function of measurement-*procedure* and/ or *-time* only. Thus, for any specified source-of-sample ($\boldsymbol{S}$), the measured estimate "${}^{S}\boldsymbol{r}_i$" should be correlated to the *unknown* true value "${}^{S}\boldsymbol{R}_i$", at best, as:

$$ {}^{S}\boldsymbol{R}_i = ({}^{S}\boldsymbol{r}_i + \Delta_S) = ({}^{S}\boldsymbol{r}_i + [{}^{(\mathbf{RAN})}\Delta_S + {}^{(\mathbf{SYS})}\Delta_S]) $$

where $\Delta_S$, ${}^{(\mathbf{RAN})}\Delta_S$ and ${}^{(\mathbf{SYS})}\Delta_S$ stand for *total*, random and *systematic* (isotopic-fractionation) errors, respectively; and where it could be a fact that "${}^{(\mathbf{SYS})}\Delta_S \gg {}^{(\mathbf{RAN})}\Delta_S$".

Similarly, any other source ($\boldsymbol{W}$) specific result can be expressed as:

$$ {}^{W}\boldsymbol{R}_i = ({}^{W}\boldsymbol{r}_i + \Delta_W) = ({}^{W}\boldsymbol{r}_i + [{}^{(\mathbf{RAN})}\Delta_W + {}^{(\mathbf{SYS})}\Delta_W]) $$

However, if $\boldsymbol{S}$ and $\boldsymbol{W}$ should represent similar isotopic sources ($|{}^{S}\boldsymbol{R}_i - {}^{W}\boldsymbol{R}_i| \to \mathbf{0}$), then the processing cum measurements of both $\boldsymbol{S}$- and $\boldsymbol{W}$-specific samples by employing *identical*

*possible experimental conditions* (***IPECs***) should cause corresponding fractionation errors to follow one another ($^{(SYS)}\Delta_S \approx {}^{(SYS)}\Delta_W$). Moreover, the possible fractionation errors $^{(SYS)}\Delta_S$ and $^{(SYS)}\Delta_W$ (for measuring any isotopic-ratio "***i***", viz. $^2H/^1H$) may differ from one another by magnitude but **never** by sign. This is the reason why certain mathematical operation on the estimates "$^S r_i$ and $^W r_i$", e.g. *computation* of their *ratio* "$y_A = (^S r_i / ^W r_i)$" should cause the difference-in-fractionation-errors "$(^{(SYS)}\Delta_S - {}^{(SYS)}\Delta_W) \equiv {}^{Add}\Delta$" *only* to turn out as the *effective* systematic error (rather, say, as an ***additional*** random measurement error). For example, if it happens that: $|^{(SYS)}\Delta_S| > |^{(SYS)}\Delta_W|$; then the error ($Đ_A$) of the estimated absolute ratio "$y_A$" should be decided as (cf. Eq. 3a, for: "$\Delta_J \equiv \Delta_S$" and "$\Delta_K \equiv \Delta_W$"):

$$Đ_A = (\Delta_S - \Delta_W) = ([^{(RAN)}\Delta_S + {}^{(SYS)}\Delta_S] - [^{(RAN)}\Delta_W + {}^{(SYS)}\Delta_W])$$

$$= ([^{(RAN)}\Delta_S + {}^{Add}\Delta] - {}^{(RAN)}\Delta_W) \qquad (3a^{/})$$

Similarly, for "$|^{(SYS)}\Delta_S| < |^{(SYS)}\Delta_W|$":

$$Đ_A = (^{(RAN)}\Delta_S - [^{(RAN)}\Delta_W + {}^{Add}\Delta]). \qquad (3a^{//})$$

Thus, even though the ***S***- and ***W***-measurement-uncertainties should be decided as: $u_S = (^{(RAN)}u_S + {}^{(SYS)}u_S)$; and: $u_W = (^{(RAN)}u_W + {}^{(SYS)}u_W)$, respectively; the ratio-uncertainty ($\varepsilon_A$, cf. Eq. 4a) should turn out to be ***less*** than the sum "$(u_S + u_W)$". That is, $\varepsilon_A$ should, although *for a case of fractionation* be ***difficult*** to be *predicted*, have a value as:

$$\varepsilon_A = (^{(RAN)}u_S + {}^{(RAN)}u_W + |^{(SYS)}u_S - {}^{(SYS)}u_W|) = (^{(RAN)}u_S + {}^{(RAN)}u_W + {}^{Add}u) \qquad (4a^{//})$$

Moreover, the basic purpose of measuring ***S*** and ***W*** by employing ***IPECs*** is to ensure that "$^{Add}u \rightarrow 0$". Thus, there should be no alternative to ascertaining "$\varepsilon_A$" as the "*lab-evaluated* **MPV** of the error $Đ_A$" (i.e.: $\varepsilon_A = {}^{Max}|Đ_A|$).

However, any lighter elemental isotopic data is acquired and/ or reported[1,2,7-15] as a *differential* ratio (viz.: $\boldsymbol{y_\delta} = [(^S\boldsymbol{r_i}/^W\boldsymbol{r_i}) - \mathbf{1}] = [\boldsymbol{y_A} - \mathbf{1}]$, rather than as the *absolute* ratio "$\boldsymbol{y_A}$"). However, should the said-practice be justified?

**3.2 Should "$f_A$" or "$f_\delta$" be the IRMS-model?**

It is pointed out above that, if only the modeled *system-functioning* (and/ or achievable *output-accuracy*) should be needed to be governed by the system-defining variables (viz. "$^S\boldsymbol{R_i}$ and $^W\boldsymbol{R_i}$") themselves, then only the "F.2" modeling (here "$\boldsymbol{Y_\delta} = ([^S\boldsymbol{R_i}/^W\boldsymbol{R_i}] - \mathbf{1}) = (\boldsymbol{Y_A} - \mathbf{1})$") has to be preferred over the "F.1" modeling (as "$\boldsymbol{Y_A} = (^S\boldsymbol{R_i}/^W\boldsymbol{R_i})$"). However, this cannot be the requirement for any analytical method, at least, for the IRMS evaluation.

Further, "$\boldsymbol{Y_A} = (^S\boldsymbol{R_i}/^W\boldsymbol{R_i})$" and "$\boldsymbol{Y_\delta} = (\boldsymbol{Y_A} - \mathbf{1})$" are shown to be so correlated (cf. e.g. Eq. 4b or 6b) that the *ratio* of corresponding *modeling-errors* (i.e. ratio of errors due to a pair of *absolute* and *differential* estimates "$\boldsymbol{y_A}$" and "$\boldsymbol{y_\delta}$", respectively) should always be ***prefixed*** as:

$$\left({Đ_\delta}/{Đ_A}\right) = \left({ε_\delta}/{ε_A}\right) = \left({[UF]_\delta}/{[UF]_A}\right) = \left|\frac{^S R_i}{^S R_i - {^W R_i}}\right| \qquad (7)$$

Eq. 7 clarifies that the knowledge of relevant "$^S\boldsymbol{R_i}$ and $^W\boldsymbol{R_i}$" *certified* materials (or even simply *theoretical* standards) should help to 'a priori' ensure whether "$\boldsymbol{Y_A} = (^S\boldsymbol{R_i}/^W\boldsymbol{R_i})$" or "$\boldsymbol{Y_\delta} = ([^S\boldsymbol{R_i}/^W\boldsymbol{R_i}] - \mathbf{1})$" be the appropriate IRMS-model. Thus, let's consider the $^2H/^1H$ certified[16] materials, IAEA-CH-7 and GISP, as the sample ***S*** and lab-reference ***W***, respectively, i.e. say that[16,17] (true): $^S\boldsymbol{R_i} = 14.013260\times10^{-5}$; and $^W\boldsymbol{R_i} = 12.62076552\times10^{-5}$. Then (cf. Eq. 7) it is predicted that:

$$({Đ_\delta}/{Đ_A}) = ({ε_\delta}/{ε_A}) = ({[UF]_\delta}/{[UF]_A}) = 10.0634 \qquad (7a)$$

Therefore:

$\text{Đ}_{\delta}$ = (10.0634 × $\text{Đ}_{A}$) (3b$^{/}$)

And/ or:

$\varepsilon_{\delta}$ = (10.0634 × $\varepsilon_{A}$) = (10.0634 × [2 × $^{G}u$]) = (20.1268 × $^{G}u$) (4b$^{///}$)

where $^{G}u$ should stand for *bias* ***corrected*** measurement-uncertainty (cf. Eq. 4a$^{/}$ or 4b$^{/}$)

It may also here be reminded that "$|^{S}R_i - {}^{W}R_i| \rightarrow \mathbf{0}$" is a requirement for IRMS evaluation. However, "$|^{S}R_i - {}^{W}R_i| \rightarrow \mathbf{0}$" should**: (i)** on the one ensure the **(**measurement-specific fractionation errors "$^{(SYS)}\Delta_S$ and $^{(SYS)}\Delta_W$" to be increasingly close to one another, and hence, in turn, the ratio-error "$\text{Đ}_{A}$" and/ or the ratio-uncertainty "$\varepsilon_{A}$" to be small; i.e.**)** estimated absolute ratio $y_A$ to be accurate; and**: (ii)** on the other cause **(**"$[UF]_{\delta} \rightarrow \infty$" and/ or "$\varepsilon_{\delta} \longrightarrow \infty$", and therefore**)** the **δ**-estimate $y_{\delta}$ to be increasingly *inaccurate*. Further, in practice (*unknown* case), "$|^{S}R_i - {}^{W}R_i|$" cannot be known beforehand. Therefore, the choice of "$Y_{\delta}$ **= ([**$^{S}R_i/{}^{W}R_i$**] – 1)**", rather than of "$Y_A$ **= (**$^{S}R_i/{}^{W}R_i$**)**", as the IRMS-model should not only mean the ***lowering*** of achievable-accuracy but also cause, at least in some cases, the evaluated data (**δ**-estimate as $y_{\delta}$, and hence the correspondingly extracted insight) to be *misleading*.

In our present *know* case, i.e. even for "$|^{S}R_i - {}^{W}R_i|$" to be somewhat *significant*, "$y_{\delta}$" is predicted to be **≈10** times more *erroneous* than "$y_A$". However, are we correct?

### 3.2.1 *Verification*

For "***S*** as IAEA-CH-7 and ***W*** as GISP", the nature of estimates ("$^{S}r_i \equiv x_J$" and "$^{W}r_i \equiv x_K$") to be expected for measurements under *varying* possible lab experimental set-ups (Nos. 1 and 2), and hence the corresponding possible *variations* of the modeled-estimates "$y_A$" and "$y_{\delta}$", are exemplified (cf. Nos. 1-5) in Table 1. However, the Expt. No. 0 **(**i.e. which shows the measurement-errors "$\Delta_S$ and $\Delta_W$", and thus the modeling errors "$\text{Đ}_{A}$ and $\text{Đ}_{\delta}$" as ***zero*)** should represent the true values, i.e.: (***true-*** $^{S}\boldsymbol{r_i} \equiv {}^{S}\boldsymbol{R_i}$) = 14.013260×10$^{-5}$ and (***true-*** $^{W}\boldsymbol{r_i} \equiv {}^{W}\boldsymbol{R_i}$) = 12.62076552×10$^{-5}$; and thus "(***true-y***$_{\boldsymbol{A}}$ ≡ $\boldsymbol{Y_A}$) = ($^{S}\boldsymbol{R_i}/^{W}\boldsymbol{R_i}$) = 1.1103336, and/ or (***true-y***$_{\boldsymbol{\delta}}$ ≡ $\boldsymbol{Y_\delta}$) ≡ ($\boldsymbol{Y_A}$ **− 1**) = 0.1103336". However, what is significant noting is that the *evaluated* ratio-of-modeling-errors, "$\boldsymbol{Đ_\delta/Đ_A}$" (cf. Table 1 for any Lab and Expt. Nos.), is the same as the *predicted* value (10.0634, cf. Eq. 7a).

However, it should be more interesting to note that the *measurement* accuracy ($\boldsymbol{u_i}$) is reflected, by Lab 1, to be reasonable (as**:** [$^{\mathbf{Max}}|\Delta_i|$ = 0.019%] ≈ 0.02%, cf. Example no. 5) and, by Lab 2, to be as worse as 0.12%. Therefore, the modeled-estimates, e.g. $\boldsymbol{y_A}^{(\mathrm{Lab1})}$ and $\boldsymbol{y_A}^{(\mathrm{Lab2})}$, should be expected (i.e. from the viewpoint of *measurement-accuracy* only, cf. Eq. 4a$^{/}$) to be 04% and 0.24% accurate, respectively. However, even $\boldsymbol{y_A}^{(\mathrm{Lab2})}$ has appeared to be 0.04% accurate, rather as erroneous as $\boldsymbol{y_A}^{(\mathrm{Lab1})}$ (because [cf. Table for any of Expt. Nos. 1-5]**:** "$|\boldsymbol{Đ_A}^{(\mathrm{Lab1})}| \leq$ 0.034%" and "$|\boldsymbol{Đ_A}^{(\mathrm{Lab2})}| \leq$ 0.036%."). Therefore, the presumable experimental conditions ("***IPECs***"), as at least those reflected by the **Lab 2** data, should be *fractionation prone*.

### 3.2.2 *Fractionation and achievable value of* $\boldsymbol{\varepsilon_d}$

It may be noted that the data ($^{S}\boldsymbol{r_i}$ and $^{W}\boldsymbol{r_i}$) by Lab 1 do not help distinguish between the possible fractionation and random errors; i.e. appear to be *unbiased*. Moreover, the ratio-error "$|\boldsymbol{Đ_A}|$" has, for the case of either Example No. 1 or 3, turned out to be *less* than even a corresponding measurement-error "$|\Delta_S|$" or "$|\Delta_W|$". This supplements the above indicated fact that the *computation of a ratio* of any two estimates (here**:** $^{S}\boldsymbol{r_i}$ and $^{W}\boldsymbol{r_i}$) should offer the possibility of partial or full cancellation (i.e. depending upon the *signs* and *magnitudes*) of the corresponding errors ($\Delta_S$ and $\Delta_W$, respectively); and thus controlling the ratio-error "$\boldsymbol{Đ_A}$". However, in the case of Example No. 2 or 4, "$|\boldsymbol{Đ_A}|$" has equaled the measurement-error-sum ($|\Delta_S|$ + $|\Delta_W|$). Therefore, the prediction as Eq. 4a$^{/}$ (i.e. *ratio-uncertainty*: $\varepsilon_A = [(\boldsymbol{u_S} + \boldsymbol{u_W}) \equiv (\boldsymbol{u_J} + \boldsymbol{u_K})] = 2\boldsymbol{u_i}$) should, at least for *unbiased* cases of measurements, be a fact. Clearly, the reason is that the different measurement-specific *random* errors "$\Delta_S \equiv \Delta_J$" and "$\Delta_W \equiv \Delta_K$" can, even *by the **sign-of-error***, differ from one another.

However, the **Lab 2** data are (at least, with reference to Lab 1) highly erroneous and/ or biased, because (cf., for illustration, the Example No. 1)**:** $(\Delta_S^{\mathbf{(Lab2)}} - \Delta_S^{\mathbf{(Lab1)}})$ = **–**0.09% and**:** $(\Delta_W^{\mathbf{(Lab2)}} - \Delta_W^{\mathbf{(Lab1)}})$ = **–**0.09**%**. That is, the *fractionation errors* $({}^{\mathbf{(SYS)}}\Delta_S$ and ${}^{\mathbf{(SYS)}}\Delta_W)$ appear to be as high as 0.09% or so. However, the measurements of ***S*** and ***W***, i.e. even by employing ***IPECs***, should *not* ensure that**:** ${}^{\mathbf{(SYS)}}\Delta_S = {}^{\mathbf{(SYS)}}\Delta_W$. In other words, the ***S*** and ***W*** specific *fractionation* and *random* errors can, in case of a real world experiment as the Example No. 1, have any combination of values as:

(i) "$\Delta_S$ = **–**0.07% = $({}^{\mathbf{(SYS)}}\Delta_S + {}^{\mathbf{(RAN)}}\Delta_S)$ = (<u>–0.09% + 0.02%</u>)"; and "$\Delta_W$ = **–**0.079% = $({}^{\mathbf{(SYS)}}\Delta_W + {}^{\mathbf{(RAN)}}\Delta_W)$ = (<u>–0.09% + 0.011%</u>)";

(ii) "$\Delta_S$ = **–**0.07% = $({}^{\mathbf{(SYS)}}\Delta_S + {}^{\mathbf{(RAN)}}\Delta_S)$ = (<u>–0.08% + 0.01%</u>)"; and "$\Delta_K$ = **–**0.079% = $({}^{\mathbf{(SYS)}}\Delta_W + {}^{\mathbf{(RAN)}}\Delta_W)$ = **[**(–0.09% + 0.011%) = (–0.08% + [**–0.01**% + 0.011%]) = $({}^{\mathbf{(SYS)}}\Delta_W + [{}^{\mathbf{\mathit{Add}}}\boldsymbol{\Delta} + {}^{\mathbf{(RAN)}}\Delta_W])$**]** = (<u>–0.08% + [**0.001**%]</u>)"; etc.

Yet, any modeled estimate ("$\boldsymbol{y_A}$" or, "$\boldsymbol{y_\delta}$") could be seen to be accountable by the theory. Thus, e.g. whether the Example No 1 should correspond to the *error-combination* no. either (i) or (ii) or some other, the ratio-error "$\boldsymbol{Đ_A}^{\text{(Expt.1)}}$" is here *predicted* (cf. Eq. 3a or Eq. 3a$^{/}$) to be 0.009%. Moreover, Table 1 (cf. for Lab 2) verifies that**:** $\boldsymbol{Đ_A}^{\text{(Expt.1)}}$ = 0.009%; and/ or that**:** $\boldsymbol{Đ_A}^{\text{(Lab2)}} = \boldsymbol{Đ_A}^{\text{(Lab1)}}$.

Thus, as shown here, the **MPV** *of modeling error* "$^{\mathbf{Max}}|\mathbf{Đ}_A|$" should be more or less *independent* of the labs (because: $^{\mathbf{Max}}|\mathbf{Đ}_A^{(\mathrm{Lab1})}|$ = 0.0342%; and: $^{\mathbf{Max}}|\mathbf{Đ}_A^{(\mathrm{Lab2})}|$ = 0.0363%). That is the ratio-uncertainty $\varepsilon_A$ should, *for a case of fractionations*, be **less** than the sum of established measurement-uncertainties "($u_S + u_W$)". In other words, *lab observed value* of the error "$^{\mathbf{Max}}|\mathbf{Đ}_A|$" should be, as indicated by Eq. 4a$^{//}$ or so, the authentic measure of "$\varepsilon_A$".

Moreover, the $^{\mathrm{Lab2}}\boldsymbol{\delta}$-estimates (with: $|\mathbf{Đ}_\delta^{(\mathrm{Lab2})}| \le$ **0.363**%) are also difficult to be distinguished from the $^{\mathrm{Lab1}}\boldsymbol{\delta}$-estimates (because: $|\mathbf{Đ}_\delta^{(\mathrm{Lab1})}| \le$ **0.342**%). This means that the measure of even the accuracy "$\varepsilon_\delta$" should be the lab-evaluated "$^{\mathbf{Max}}|\mathbf{Đ}_\delta|$".

Therefore, if the aim of modeling should be to have a means for ***simply*** cancelling the possible ***S***- and ***W***-specific fractionation errors, i.e. ***not*** really *for achieving* the best possible accuracy in the desired (modeled) result, then "$\boldsymbol{Y_\delta = ([{}^{S}R_i/{}^{W}R_i] - 1) = (Y_A - 1)}$" should be equally as suitable a model as "$\boldsymbol{Y_A = ({}^{S}R_i/{}^{W}R_i)}$" for the case of IRMS. However, it is demonstrated that (cf. Table 1), and/ or explained why (cf. Eq. 4b$^{///}$ or Eq. 7a or so), the usual IRMS-estimate "$\boldsymbol{y_\delta}$" should be more *erroneous* than the corresponding absolute ratio "$\boldsymbol{y_A}$".

### 3.3 Result: $^{S/D}\delta_i$-estimate ($z_\delta$) or absolute estimate ($z_A$, or $^{S}r_i$)?

It is pointed out above that, for enabling the comparison between different possible lab-results, any lab (***W***) specific $^{S/W}\boldsymbol{\delta_i}$-estimate ($\boldsymbol{y_\delta}$) should be translated into the corresponding recommended[10,11] standard (***D***) specific $^{S/D}\boldsymbol{\delta_i}$-estimate ($\boldsymbol{z_\delta}$); i.e. reference-scale-transformation "$\boldsymbol{W \rightarrow D}$" is a general requirement Thus, by any IRMS evaluation, it should mean that:

$$\boldsymbol{Z_\delta} = [\boldsymbol{h_\delta(Y_\delta) \equiv h_\delta(f_\delta(Y_A))}] \quad (8)$$

Or, in terms of estimates:

$$\boldsymbol{(z_\delta \pm {}^{Z}\varepsilon_\delta)} = [\boldsymbol{h_\delta([y_\delta \pm \varepsilon_\delta]) \equiv h_\delta(f_\delta(y_A \pm \varepsilon_A))}] \quad (8^{/})$$

where (the desired result, i.e. *scale converted* **δ**-ratio)**:** $\boldsymbol{z_\delta}$ = ([$^{S}\boldsymbol{r_i}$/$^{D}\boldsymbol{R_i}$] − **1**) = ($\boldsymbol{z_A}$ − **1**); and $^{Z}\boldsymbol{\varepsilon_\delta}$ is the corresponding (i.e. **δ**-*scale conversion*) uncertainty.

Further, let's refer to the uncertainty of the *scale converted* ***absolute*** ratio ($\boldsymbol{z_A}$) as "$^{Z}\boldsymbol{\varepsilon_A}$". However, it is (in terms of "***S/W***" estimates) shown above that "$\boldsymbol{\varepsilon_\delta} > \boldsymbol{\varepsilon_A}$". Then, shouldn't it be true that "$^{Z}\boldsymbol{\varepsilon_\delta} > {}^{Z}\boldsymbol{\varepsilon_A}$"?

Therefore, it is felt imperative to examine whether the *IRMS principle* (cf. Eq. 8) "$\boldsymbol{Z_\delta} = \boldsymbol{h_\delta}(\boldsymbol{Y_\delta})$" should itself be worth reformulating as "$\boldsymbol{Z_A} = \boldsymbol{h_A}(\boldsymbol{Y_A})$" or so.

As**:** $\boldsymbol{Y_\delta}$ = ([$^{S}\boldsymbol{R_i}$/$^{W}\boldsymbol{R_i}$] − **1**) = ($\boldsymbol{Y_A}$ − **1**), and**:** $\boldsymbol{Z_\delta}$ = ([$^{S}\boldsymbol{R_i}$/$^{D}\boldsymbol{R_i}$] − **1**) = ($\boldsymbol{Z_A}$ − **1**); the evaluation of a result as "$\boldsymbol{z_\delta}$ (or even $\boldsymbol{z_A}$, or $^{S}\boldsymbol{r_i}$)" should require[2,13-15,18] "***W***" to be a *calibrated* reference material. Else, [2,13-15,18] certain ***other*** *calibrated* materials (we say,[17,19] ***auxiliary*** reference-standards "***Ai***, with**:** ***i*** = 1, 2 **…**") should also, i.e. in addition to the usual sample ***S*** and by employing the investigating lab-established "***IPECs***", be measured.

It may further be pointed out that a *true value* (i.e. any *variable*, e.g. "$\boldsymbol{Z_\delta}$ or, $^{S}\boldsymbol{R_i}$" or so) ***cannot*** be method-specific. However, any *estimate* as "$\boldsymbol{z_\delta}$" (i.e. achievable output-accuracy "$^{Z}\boldsymbol{\varepsilon_\delta}$") should be method-dependent. Thus, simply for *distinguishing* between the different possible characteristics of different typical scale-conversion-methods, we may refer to the employing of "calibrated ***W***", "only one "***Ai***" and "two different ***Ai***-standards" as the **Mtd-1**, **Mtd-2** and **Mtd-3**, respectively.

### 3.3.1 **Mtd-1:** ***W** calibrated method of scale conversion*

♦ 1. <u>Scale conversion of *ratio-of-ratios*</u> ($\boldsymbol{Y_A} \rightarrow \boldsymbol{Z_A}$)

The expression "$\boldsymbol{Z_A}$ = ($^{S}\boldsymbol{R_i}$/$^{D}\boldsymbol{R_i}$)" should itself help derive the formula "$\boldsymbol{Z_A} = \boldsymbol{h_A}(\boldsymbol{Y_A})$":

$$Z_A = \left(\frac{{}^{S}R_i}{{}^{D}R_i}\right) = \left(\frac{{}^{S}R_i}{{}^{W}R_i} \times \frac{{}^{W}R_i}{{}^{D}R_i}\right) = (Y_A \times C_A) = (Y_A \times [C_\delta + 1]) \tag{9}$$

Therefore, in terms of estimates, Eq. 9 could be rewritten as:

$$(z_A^{(\text{Mtd-1})} \pm {}^{Z}\varepsilon_A^{(\text{Mtd-1})}) = \mathbf{(}[y_A \pm \varepsilon_A] \times [C_\delta + 1]) \qquad (9')$$

where $\boldsymbol{C_\delta}$ stands for the *known* "***W*** vs. ***D***" isotopic calibration constant (i.e.**:** $\boldsymbol{C_\delta} = [\mathbf{(}{}^{W}\boldsymbol{R_i}/{}^{D}\boldsymbol{R_i}) - \mathbf{1}]$ $= [\boldsymbol{C_A} - \mathbf{1}]$); e.g. here,[16] i.e. for "***W*** as GISP" and "***D*** as VSMOW" hydrogen**:** $\boldsymbol{C_\delta} = -0.18973$.

However, Eq. 9 should, like Eq. 1, belong to the F.1 family, i.e. "$\boldsymbol{Z_A}$" could be shown to be ***equally*** as sensitive as "$\boldsymbol{Y_A}$" towards a possible measurement-variation, and thus (uncertainty-factor, cf. Eq. 5/ 6)**:** ${}^{Z}[\boldsymbol{UF}]_A^{(\mathbf{Mtd\text{-}1})} = \left| {}^{A}\boldsymbol{M}_{S/W}^{S/D} \right| = 1$. Therefore, the uncertainty, ${}^{Z}\boldsymbol{\varepsilon_A}^{(\mathbf{Mtd\text{-}1})}$, of determining the *scale converted absolute* value ($\boldsymbol{z_A}^{(\mathbf{Mtd\text{-}1})}$) should be decided[6] as (cf. also Eq. 4):

$${}^{Z}\boldsymbol{\varepsilon_A}^{(\mathbf{Mtd\text{-}1})} = ({}^{Z}[UF]_A^{(\text{Mtd-1})} \times \varepsilon_A) = \boldsymbol{\varepsilon_A} \qquad (4c)$$

That is, **Mtd-1** specific (***S/D***) ratio-of-ratios ($\boldsymbol{z_A}^{(\mathbf{Mtd\text{-}1})}$) is predicted to be *as accurate as* the **lab**-estimated, i.e. "***S/W***", ratio-of-ratios ($\boldsymbol{y_A}$).

Further, if the measurement conditions "***IPECs***" should (like the Lab 1 in Table 1) help suppress the fractionation to the effects that (the ***S***- and ***W****-measurement-uncertainties*)**:** $\boldsymbol{u_S} = {}^{(\mathbf{RAN})}\boldsymbol{u_S}$, and $\boldsymbol{u_W} = {}^{(\mathbf{RAN})}\boldsymbol{u_W}$ (respectively); and if also**:** $\boldsymbol{u_S} = \boldsymbol{u_W} = {}^{G}\boldsymbol{u}$; then "${}^{Z}\varepsilon_A$" can be expressed as (cf. Eq. 4a′):

$${}^{Z}\boldsymbol{\varepsilon_A}^{(\mathbf{Mtd\text{-}1})} = \boldsymbol{\varepsilon_A} = (u_S + u_W) = (2 \times {}^{G}\boldsymbol{u}) \qquad (4c')$$

♦1.1. <u>Evaluation of sample isotopic abundance ratio</u> ($\boldsymbol{Z_A} \rightarrow {}^{S}\boldsymbol{R_i}$)

As ${}^{D}\boldsymbol{R_i}$ should ever be known, the estimate "${}^{S}\boldsymbol{r_i}$" can also always be computed:

$${}^{S}R_i = ({}^{D}R_i \times Z_A) \qquad (10)$$

And hence:

$$({}^{S}r_i \pm {}^{S}\varepsilon_i) = ({}^{D}R_i \times [z_A^{(\text{Mtd-1})} \pm {}^{Z}\varepsilon_A^{(\text{Mtd-1})}]) \qquad (10')$$

Eq. 10 could also be shown to belong to the F.1 model-family; i.e. (cf. Eq. 5/ 6)**:** $[\boldsymbol{UF}]_{\boldsymbol{Z_A}}^{\boldsymbol{S_{R_i}}}$ = $\left|\boldsymbol{M}_{\boldsymbol{Z_A}}^{\boldsymbol{S_{R_i}}}\right|$ = 1. Thus the uncertainty $^{S}\boldsymbol{\varepsilon_i}$ should be fixed as (cf. Eq. 4 and/ or Eq. 4c):

$$^{S}\boldsymbol{\varepsilon_i} = ([UF]_{Z_A}^{S_{R_i}} \times {}^{Z}\varepsilon_A{}^{(\text{Mtd-1})}) = {}^{Z}\boldsymbol{\varepsilon_A}{}^{(\textbf{Mtd-1})} = \boldsymbol{\varepsilon_A} \qquad (4d)$$

And, for the possible *bias* (fractionation) ***free*** cases (as those referred to by Eq. 4c$^{/}$):

$$^{S}\boldsymbol{\varepsilon_i} = {}^{Z}\boldsymbol{\varepsilon_A}{}^{(\textbf{Mtd-1})} = \boldsymbol{\varepsilon_A} = (u_S + u_W) = (2 \times {}^{G}\boldsymbol{u}) \qquad (4d^{/})$$

That is, "$^{S}\boldsymbol{r_i}$" should turn out ***equally*** as accurate as the corresponding "***S/D***" ratio-of-ratios (here, $\boldsymbol{z_A}^{(\textbf{Mtd-1})}$), and/ or as the "***S/W***" (i.e. ***lab****-estimated*) ratio-of-ratios ($\boldsymbol{y_A}$).

♦2. <u>**δ**-Scale conversion</u> ($\boldsymbol{Y_\delta} \rightarrow \boldsymbol{Z_\delta}$)

Eq. 8 (i.e. "$\boldsymbol{Z_\delta = h_\delta(Y_\delta)}$") should, like Eq. 9, be arrived as follows[17,19]:

$$Z_\delta = \left(\frac{{}^{S}R_i}{{}^{D}R_i} - 1\right) = \left(\frac{{}^{S}R_i}{{}^{W}R_i} \times \frac{{}^{W}R_i}{{}^{D}R_i} - 1\right) = ([Y_\delta + 1] \times [C_\delta + 1] - 1) \qquad (11)$$

Thus Eq. 8$^{/}$ (i.e. "$[\boldsymbol{z_\delta} \pm {}^{Z}\boldsymbol{\varepsilon_\delta}] = \boldsymbol{h_\delta(y_\delta \pm \varepsilon_\delta)}$") should take the description as:

$$(z_\delta{}^{(\text{Mtd-1})} \pm {}^{Z}\varepsilon_\delta{}^{(\text{Mtd-1})}) = \boldsymbol{(}[(y_\delta \pm \varepsilon_\delta) + 1] \times [C_\delta + 1] - 1) \qquad (11^{/})$$

However, it is already shown elsewhere[17,19] that Eq. 11 belongs to the F.2 family; i.e. the variation of "$\boldsymbol{Z_\delta}$" as a function of "$\boldsymbol{Y_\delta}$" should be fixed as[17] (cf. Eq. 5/ 6)**:** $^{Z}[\boldsymbol{UF}]_{\boldsymbol{\delta}}{}^{(\textbf{Mtd-1})}$ = $\left|{}^{\delta}\boldsymbol{M}_{\boldsymbol{S/W}}^{\boldsymbol{S/D}}\right|$ = $\left|\boldsymbol{(^{S}R_i - {}^{W}R_i)/(^{S}R_i - {}^{D}R_i)}\right|$; i.e. the uncertainty $^{Z}\boldsymbol{\varepsilon_\delta}{}^{(\textbf{Mtd-1})}$ (cf. Eq. 4 or 4b or so) as**:**

$$^{Z}\varepsilon_\delta{}^{(\text{Mtd-1})} = (^{Z}[UF]_\delta{}^{(\text{Mtd-1})} \times \varepsilon_\delta) = \left(\left|\frac{{}^{S}R_i - {}^{W}R_i}{{}^{S}R_i - {}^{D}R_i}\right| \times \varepsilon_\delta\right) \qquad (4E)$$

Further, "***D***, ***S*** and ***W***" should all represent similar (viz. natural) isotopic materials. Therefore, "$^{Z}[\boldsymbol{UF}]_{\boldsymbol{\delta}}{}^{(\textbf{Mtd-1})}$" should be close to unity. However, for illustration, say that "$\boldsymbol{(^{D}R_i/^{W}R_i)}$ **<** **1**". Then, the **δ**-*scale conversion* uncertainty "$^{Z}\boldsymbol{\varepsilon_\delta}{}^{(\textbf{Mtd-1})}$" should be somewhat **higher** than the corresponding IRMS (i.e. **δ**-*measurement*) uncertainty "$\boldsymbol{\varepsilon_\delta}$". Otherwise (i.e. for**:** $[\boldsymbol{^{D}R_i/^{W}R_i}]$ **>** **1**), the ${}^{S/D}\boldsymbol{\delta}$-estimate ($\boldsymbol{z_\delta}^{\textbf{(Mtd-1)}}$) should turn out more *accurate* than the corresponding lab, i.e. ${}^{S/W}\boldsymbol{\delta}$-, estimate ($\boldsymbol{y_\delta}$). Thus, as[16]**:** ${}^{D}\boldsymbol{R_i}$ = 15.576×10$^{-5}$, the present known case should imply that**:** ${}^{Z}[\boldsymbol{UF}]_{\boldsymbol{\delta}}^{\textbf{(Mtd-1)}}$ = 0.891; and/ or that**:** ${}^{Z}\boldsymbol{\varepsilon_\delta}^{\textbf{(Mtd-1)}}$ = (0.891 × $\boldsymbol{\varepsilon_\delta}$).

However, as (cf. Eqs. 4b$^{///}$)**:** $\varepsilon_\delta$ = (10.0634 × $\boldsymbol{\varepsilon_A}$); and (cf. Eq. 4c$^{/}$)**:** ${}^{Z}\boldsymbol{\varepsilon_A}^{\textbf{(Mtd-1)}}$ = $\boldsymbol{\varepsilon_A}$ = ($u_S$ + $u_W$) = (2 × ${}^{G}\boldsymbol{u}$); the **δ**-scale conversion uncertainty " ${}^{Z}\boldsymbol{\varepsilon_\delta}^{\textbf{(Mtd-1)}}$" could be re-expressed as:

$${}^{Z}\boldsymbol{\varepsilon_\delta}^{\textbf{(Mtd-1)}} = (0.891 \times [10.0634 \times \boldsymbol{\varepsilon_A}]) = (8.967 \times \boldsymbol{\varepsilon_A}) = (8.967 \times {}^{Z}\varepsilon_A^{\textbf{(Mtd-1)}})$$

$$= (8.967 \times [u_S + u_W]) = (8.967 \times [\mathbf{2} \times {}^{G}\boldsymbol{u}]) = (17.934 \times {}^{G}\boldsymbol{u}) \qquad (4E^{/})$$

where ${}^{G}\boldsymbol{u}$ should stand for *bias free* measurement-uncertainty**:** ${}^{G}\boldsymbol{u}$ **=** **(**$\boldsymbol{u_S}$ ≡ ${}^{\textbf{(RAN)}}\boldsymbol{u_S}$) = ($\boldsymbol{u_W}$ ≡ ${}^{\textbf{(RAN)}}\boldsymbol{u_W}$).

<u>Alternative process</u> ($\boldsymbol{Z_A} \rightarrow \boldsymbol{Z_\delta}$)

The result ($\boldsymbol{z_\delta}$) can also be evaluated from "$\boldsymbol{z_A}$"; i.e. Eq. 11 should be equivalent to:

$$Z_\delta = (Z_A - 1) \qquad (11a)$$

And, therefore:

$$(z_\delta^{\text{(Mtd-1)}} \pm {}^{Z}\varepsilon_\delta^{\text{(Mtd-1)}}) = ([z_A^{\text{(Mtd-1)}} \pm {}^{Z}\varepsilon_A^{\text{(Mtd-1)}}] - 1) \qquad (11a^{/})$$

The "$\boldsymbol{Z_\delta}$ versus $\boldsymbol{Z_A}$" variation-rate could be shown to be fixed as (cf. Eq. 5/ 6)**:** ${}^{Z}[\boldsymbol{UF}]_{\boldsymbol{A}}^{\boldsymbol{\delta}}$ = $\left|{}^{Z}\boldsymbol{M}_{\boldsymbol{A}}^{\boldsymbol{\delta}}\right|$ = $\left|(\boldsymbol{Z_A}/[\boldsymbol{Z_A} - \mathbf{1})\right|$ = $\left|({}^{S}\boldsymbol{R_i}/[{}^{S}\boldsymbol{R_i} - {}^{D}\boldsymbol{R_i}])\right|$ = 8.9671; and thus (cf. Eq. 4, and also Eq. **4E$^{/}$**)**:**

$${}^{Z}\boldsymbol{\varepsilon_\delta}^{\textbf{(Mtd-1)}} = ({}^{Z}[\boldsymbol{UF}]_{\boldsymbol{A}}^{\boldsymbol{\delta}} \times {}^{Z}\varepsilon_A^{\textbf{(Mtd-1)}}) = (8.967 \times {}^{Z}\varepsilon_A^{\textbf{(Mtd-1)}}) = (8.967 \times \boldsymbol{\varepsilon_A}) \qquad (4E^{//})$$

Eq. 4E$^{//}$ supplements the prediction (cf. the context of Eq. 2 and Eq. 2$^{/}$) that a result should ***not*** vary for varying the evaluation-path *only*. That is, irrespective of whether one should go by Eq. 11 or Eq. 11a, the ${}^{S/D}\boldsymbol{\delta}$-estimate $\boldsymbol{z_\delta}^{\textbf{(Mtd-1)}}$ is predicted to turn out more *erroneous* (here, **≈9** times) than any corresponding ***absolute*** *ratio* (as either "***S/D***" estimate $\boldsymbol{z_A}^{\textbf{(Mtd-1)}}$, or lab-"***S/W***"-estimate $\boldsymbol{y_A}$).

♦2.1. Should the estimates "$^{S}\boldsymbol{r_i} = \boldsymbol{f_S}(\boldsymbol{z_A})$" and "$^{S}\boldsymbol{r_i} = \boldsymbol{f_S}(\boldsymbol{z_\delta})$" be different?

The computation of "$^{S}\boldsymbol{r_i}$" from "$\boldsymbol{z_\delta}$" should be an equally simple task as Eq. 10:

$$^{S}R_i = (^{D}R_i \times [Z_\delta + 1]) \qquad (10a)$$

That is:

$$(^{S}r_i \pm {}^{S}\varepsilon_i) = (^{D}R_i \times [(z_\delta{}^{(\text{Mtd-1})} \pm {}^{Z}\varepsilon_\delta{}^{(\text{Mtd-1})}) + 1]) \qquad (10a')$$

The *uncertainty-factor* of evaluating "$^{S}\boldsymbol{R_i}$" from "$\boldsymbol{Z_\delta}$" could be shown to be decided as[17]**:** $[\boldsymbol{UF}]_{\boldsymbol{Z_\delta}}^{\boldsymbol{^{S}R_i}} = \left|\boldsymbol{M}_{\boldsymbol{Z_\delta}}^{\boldsymbol{^{S}R_i}}\right| = \left|(\boldsymbol{Z_\delta/[Z_\delta + 1}])\right| = \left|(^{S}\boldsymbol{R_i} - {}^{D}\boldsymbol{R_i})/^{S}\boldsymbol{R_i}\right|$;. i.e. ("$[\boldsymbol{UF}]_{\boldsymbol{Z_\delta}}^{\boldsymbol{^{S}R_i}}$ **< 1**" and in the present known case)**:** $[\boldsymbol{UF}]_{\boldsymbol{Z_\delta}}^{\boldsymbol{^{S}R_i}} = (\mathbf{1}/^{Z}[\boldsymbol{UF}]_{\boldsymbol{A}}^{\boldsymbol{\delta}}) = (\mathbf{1/8.9671})$; and thus (cf. Eq. 4 or so) the uncertainty $^{S}\boldsymbol{\varepsilon_i}$**:**

$$^{S}\boldsymbol{\varepsilon_i} = ([\boldsymbol{UF}]_{\boldsymbol{Z_\delta}}^{\boldsymbol{^{S}R_i}} \times {}^{Z}\boldsymbol{\varepsilon_\delta}^{(\textbf{Mtd-1})}) = (^{Z}\boldsymbol{\varepsilon_\delta}^{(\textbf{Mtd-1})}/\mathbf{8.9671}) \qquad (4d'')$$

The finding here (cf. Eq. 4d′′) is really in corroboration with the previous report[17,19] that estimated *sample ratios* ($^{S}\boldsymbol{r_i}^{(\text{Lab1})}$, $^{S}\boldsymbol{r_i}^{(\text{Lab2})}$ **…**) should ***better*** represent the sample (***S***) and be *more closely* intercomparable than the corresponding **δ**-*estimates* ($\boldsymbol{z_\delta}^{(\text{Lab1})}$, $\boldsymbol{z_\delta}^{(\text{Lab2})}$ **…**).

Moreover, as (cf. Eq. 4E′′)**:** $^{Z}\boldsymbol{\varepsilon_\delta}^{(\textbf{Mtd-1})} = (8.967 \times {}^{Z}\varepsilon_{\boldsymbol{A}}{}^{(\textbf{Mtd-1})})$; Eq. 4d′′ should be equivalent to Eq. 4d′. That is, irrespective of whether the path chosen is Eq. 10 or Eq. 10a, the **Mtd-1** specific uncertainty ($^{S}\boldsymbol{\varepsilon_i}^{(\textbf{Mtd-1})}$) should be the *one and the same*:

$$^{S}\boldsymbol{\varepsilon_i}^{(\textbf{Mtd-1})} = (^{Z}\boldsymbol{\varepsilon_\delta}^{(\textbf{Mtd-1})}/8.9671) = {}^{Z}\varepsilon_{\boldsymbol{A}}{}^{(\textbf{Mtd-1})} = \boldsymbol{\varepsilon_A} = (u_S + u_W) = (2 \times {}^{G}\boldsymbol{u}) \qquad (4d''')$$

Eq. 4d′′′ further emphasizes the point that a mere change of evaluation path should not cause the desired output (here**:** $^{S}\boldsymbol{r_i}$) to be different.

However, the important finding is that the "***differential***-to-***absolute*** *uncertainty-ratio*" as either "($^{Z}\boldsymbol{\varepsilon_\delta}/^{Z}\boldsymbol{\varepsilon_A}$) ≡ ($^{S/D}\boldsymbol{\varepsilon_\delta}/^{S/D}\boldsymbol{\varepsilon_A}$)" or "($^{Z}\boldsymbol{\varepsilon_\delta}/\boldsymbol{\varepsilon_A}$) ≡ ($^{S/D}\boldsymbol{\varepsilon_\delta}/^{S/W}\boldsymbol{\varepsilon_A}$)" or "even ($^{Z}\boldsymbol{\varepsilon_\delta}/^{S}\boldsymbol{\varepsilon_i}$) ≡ ($^{S/D}\boldsymbol{\varepsilon_\delta}/^{S}\boldsymbol{\varepsilon_i}$)" should be **>1**; and/ or equal to the factor as "$\left|(^{S}\boldsymbol{R_i}/[^{S}\boldsymbol{R_i} - {}^{D}\boldsymbol{R_i}])\right|$". That is the ***ratio*** of possible *errors* **"**in any usual **IRMS** result (i.e. $^{S/D}$**δ**-estimate $\boldsymbol{z_\delta}$)**"** and **"**in *any* corresponding **absolute** estimate (as either "***S/D***" ratio-of-ratios $\boldsymbol{z_A}$ or "***S/W***" ratio-of-ratios $\boldsymbol{y_A}$ or even sample isotopic ratio $^{S}\boldsymbol{r_i}$)**"** should, depending ***only*** *on the difference in* isotopic composition (IC) between ***S*** and ***D***, be ***prefixed*** as >**1**; e.g. (here, in the known case [see also Eq. 7a]):

$$\frac{^{Z}Đ_{\delta}}{^{Z}Đ_{A}} = \frac{^{Z}Đ_{\delta}}{Đ_{A}} = \frac{^{Z}Đ_{\delta}}{^{S}Đ_{i}} = \left| \frac{^{S}R_{i}}{^{S}R_{i} - {^{D}R_{i}}} \right| = 8.9671 \qquad (7b)$$

Moreover, all the different "***S/W***" estimates ($\boldsymbol{y_A}$ and $\boldsymbol{y_\delta}$) in Table 1 are translated into the corresponding**: (i)** "***S/D***" estimates ($\boldsymbol{z_A}$ and $\boldsymbol{z_\delta}$, respectively) and**: (ii)** also $^{S}\boldsymbol{R_i}$*-values*; and furnished in Table 2, which shows that the *estimated* error-ratios (cf. columns 7 and 9) are the same as *predicted* (cf. Eq. 7b). Thus, Table 2 confirms the finding that the ***results*** *of the evaluations as* "(Eq. 9)**:** $\boldsymbol{Z_A = h_A(Y_A)}$", "(Eqs. 9-10)**:** $\boldsymbol{^{S}R_i = h_S(Z_A) = h_S(h_A(Y_A))}$" and even "(Eqs. 10a-11)**:** $\boldsymbol{^{S}R_i = h_S(Z_\delta) = h_S(h_\delta(Y_\delta))}$" should turn out **equivalent** (i.e. *equally* well represent the sample-source ***S***), and be ***more*** accurate (i.e. be better representatives) than the ***usual*** *IRMS result* to be obtained as (Eq. 11)**:** $\boldsymbol{Z_\delta = h_\delta(Y_\delta)}$.

### 3.3.2 *Scale conversion with the aid of **Ai**-standards*

It is indicated above that the employing of even a single ***Ai***-standard (i.e. **Mtd-2:** $\boldsymbol{Y_\delta \xrightarrow{A1} Z_\delta \rightarrow {^{S}R_i}}$) should require an ***Ai***-measurement, i.e. cause any desired result ($\boldsymbol{z_\delta}$ and, in turn, $^{S}\boldsymbol{r_i}$) to be subject to an *additional* source of error. Thus, even though the **Mtd-3** ($\boldsymbol{Y_\delta \xrightarrow{A1,A2} Z_\delta}$) is believed [11,14,15,18] to ensure "$\boldsymbol{z_\delta}$" to be more accurate than that to be obtained by the **Mtd-2**; it has already been clarified elsewhere[17,19] that "$\boldsymbol{Y_\delta \xrightarrow{A1,A2} Z_\delta \rightarrow {^{S}R_i}}$"; "$\boldsymbol{Y_\delta \xrightarrow{A1\ (or\ A2)} Z_\delta \rightarrow {^{S}R_i}}$" and "$\boldsymbol{Y_\delta \dashrightarrow Z_\delta \rightarrow {^{S}R_i}}$" (i.e. **Mtd-3**, **Mtd-2** and **Mtd-1**) should yield the ***least***, *moderate* and most accurate results, respectively. Yet, we may verify, below, the generality of the finding[17,19] in terms of *ratio-of-ratios* (i.e.**:** $\boldsymbol{Y_A \xrightarrow{Ai(s)} Z_A \rightarrow {^{S}R_i}}$) even.

However, it may here be reminded that there is a finite possibility[3] for any *multivariable* result (e.g. “$y_A$” or “$y_\delta$” in Table 1) to turn out to be **100%** accurate. Thus, e.g. Eq. 3a or 3a$^{/}$ clarifies that the estimate “$y_A$” should, even in a case where the corresponding individual measurement-errors be *non-zero* but equal to one another (i.e. even when: $[\Delta_S = \Delta_W] \neq 0$), represent the true value “$Y_A$” (i.e. “$Đ_A = 0$”, and hence “$Đ_\delta = 0$”, should be true). However, measurement-errors can ***never*** be preset (i.e. “$Đ_\delta = 0$” or so cannot be achieved). Thus, “uncertainty” should not be confused with the triviality as “$Đ_A = 0$” or “$Đ_\delta = 0$” or so.

We should also keep in mind that, although any measurement-uncertainty as “$u_S$” should represent the “**MPVs**” of both random and fractionation errors: $u_S = (|{}^{(RAN)}\Delta_S| + |{}^{(SYS)}\Delta_S|)$; the *ratio-error* as “$Đ_A$” should practically be governed (cf. Eq. 3a$^{/}$ or 3a$^{//}$) by the *random* measurement-errors (as ${}^{(RAN)}\Delta_S$ and ${}^{(RAN)}\Delta_W$) only. Thus the ratio-uncertainty ($\varepsilon_A$), although needs to be experimentally established under the investigating lab’s “***IPECs***” (cf. Eq. 4a$^{//}$), is expressed below as: $\varepsilon_A \approx ({}^{(RAN)}u_S + {}^{(RAN)}u_W)$. Moreover, the formula “$\varepsilon_A = 2^G u$ (cf. Eq. 4a$^{/}$)” should in the present context refer to “${}^G u$” as the fractionation *corrected* measurement uncertainty, and signify that “${}^G u = {}^{(RAN)}u_S = {}^{(RAN)}u_W$”; or (if applicable) “${}^G u = {}^{(RAN)}u_{A1} = {}^{(RAN)}u_W$”; or “${}^G u = {}^{(RAN)}u_{A2} = {}^{(RAN)}u_W$”.

3.3.2.1 **Mtd-2:** use of single ***Ai***-standard (***A*1**)

•1. <u>Scale conversion of *ratio-of-ratios*</u> ($Y_A \xrightarrow{A1} Z_A$)

Let’s, like “$Y_A$”, denote the *auxiliary* variable as “$Y1_A$”, i.e.: $Y1_A = ({}^{A1}R_i/{}^{W}R_i)$. Then the required formula “$Z_A = f_A(Y_A, Y1_A)$” can, like Eq. 9, be derived as follows:

$$Z_A = \left(\frac{{}^S R_i}{{}^D R_i}\right) = \left(\frac{{}^S R_i}{{}^W R_i} \times \frac{{}^W R_i}{{}^{A1} R_i} \times \frac{{}^{A1} R_i}{{}^D R_i}\right) = \left(\frac{Y_A \times C1_A}{Y1_A}\right) = \left(\frac{Y_A \times [C1_\delta + 1]}{Y1_A}\right) \qquad (9a)$$

That is, in terms of estimates:

$(z_A{}^{(\text{Mtd-2})} \pm {}^{Z}\varepsilon_A{}^{(\text{Mtd-2})}) = ([y_A \pm \varepsilon_A] \times [C1_\delta + 1])/(y1_A \pm \varepsilon 1_A)$ (9a$^{/}$)

where $\boldsymbol{C1_\delta}$ is the *known* **$A1$** vs. **$D$** calibration constant (i.e.: $\boldsymbol{C1_\delta} = [({}^{A1}R_i/{}^{D}R_i) - 1] = [C1_A - 1]$); and "$\boldsymbol{\varepsilon_A}$ and $\boldsymbol{\varepsilon 1_A}$" represent the *lab-established* uncertainties of the estimates "$\boldsymbol{y_A}$ and $\boldsymbol{y1_A}$", respectively; i.e. (cf. Eq. 4a$^{//}$): $\boldsymbol{\varepsilon_A} \approx ({}^{(\text{RAN})}\boldsymbol{u_S} + {}^{(\text{RAN})}\boldsymbol{u_W})$, or (cf. Eq. 4a$^{/}$): $\boldsymbol{\varepsilon_A} = ([\boldsymbol{UF}]_A \times {}^{G}\boldsymbol{u}) = 2{}^{G}\boldsymbol{u}$; and thus: $\boldsymbol{\varepsilon 1_A} \approx ({}^{(\text{RAN})}\boldsymbol{u_{A1}} + {}^{(\text{RAN})}\boldsymbol{u_W})$, and/ or: $\boldsymbol{\varepsilon 1_A} = ({}^{Y1}[\boldsymbol{UF}]_A \times {}^{G}\boldsymbol{u}) = 2{}^{G}\boldsymbol{u}$.

However, Eq. 9a could be shown to belong, like Eq. 9, to the F.1 family. That is, even Eq. 9a specific individual variation-rates should be *invariable* (cf. Eq. 5): $\boldsymbol{M_{Y_A}^{Z_A} = 1}$, and: $\boldsymbol{M_{Y1_A}^{Z_A} = -1}$. Therefore, the uncertainty "${}^{Z}\boldsymbol{\varepsilon_A}{}^{(\text{Mtd-2})}$" should be decided as (cf. Eq. 4):

$${}^{Z}\varepsilon_A{}^{(\text{Mtd-2})} = \sum_{i=1}^{2} \left(\left|M_i^{Z_A}\right| \times \varepsilon_i\right) = \left[\left(\left|M_{Y_A}^{Z_A}\right| \times \varepsilon_A\right) + \left(\left|M_{Y1_A}^{Z_A}\right| \times \varepsilon 1_A\right)\right] = (\varepsilon_A + \varepsilon 1_A)$$

$$\approx ({}^{(\text{RAN})}\boldsymbol{u_S} + {}^{(\text{RAN})}u_W) + ({}^{(\text{RAN})}\boldsymbol{u_{A1}} + {}^{(\text{RAN})}u_W) \qquad (4f)$$

Or (recollecting that [cf. e.g. Eq. 4d$^{///}$]: ${}^{Z}\boldsymbol{\varepsilon_A}{}^{(\text{Mtd-1})} = \boldsymbol{\varepsilon_A} = {}^{S}\boldsymbol{\varepsilon_i}{}^{(\text{Mtd-1})} = \boldsymbol{2{}^{G}u}$):

$${}^{Z}\boldsymbol{\varepsilon_A}{}^{(\text{Mtd-2})} = (\varepsilon_A + \varepsilon 1_A) = ({}^{Z}\boldsymbol{\varepsilon_A}{}^{(\text{Mtd-1})} + \varepsilon 1_A) \qquad (4f^{/})$$

Or (in terms of "${}^{G}\boldsymbol{u}$"):

$${}^{Z}\boldsymbol{\varepsilon_A}{}^{(\text{Mtd-2})} = (\boldsymbol{\varepsilon_A} + \boldsymbol{\varepsilon 1_A}) = [([\boldsymbol{UF}]_A \times {}^{G}u) + ({}^{Y1}[\boldsymbol{UF}]_A \times {}^{G}u)] = [([\boldsymbol{UF}]_A + {}^{Y1}[\boldsymbol{UF}]_A) \times {}^{G}u] =$$

$$([\boldsymbol{2} + \boldsymbol{2}] \times {}^{G}u) = ({}^{Z}[\boldsymbol{UF}]_A{}^{(\text{Mtd-2})} \times {}^{G}u) = (\boldsymbol{2} \times [\boldsymbol{2} \times {}^{G}\boldsymbol{u}]) = (\boldsymbol{2} \times {}^{Z}\boldsymbol{\varepsilon_A}{}^{(\text{Mtd-1})}) \qquad (4f^{//})$$

Thus, even in terms of *ratio-of-ratios*, the **Mtd-1** (i.e. "$\boldsymbol{Z_A = h_A(Y_A)}$") should yield more accurate result than the **Mtd-2** ("$\boldsymbol{Z_A = f_A(Y_A, Y1_A)}$").

•1.1. <u>Evaluation of ${}^{S}\boldsymbol{R_i}$-value</u> ($\boldsymbol{z_A}{}^{(\text{Mtd-2})} \rightarrow {}^{S}\boldsymbol{r_i}{}^{(\text{Mtd-2})}$)

The *formula* "${}^{S}\boldsymbol{R_i} = \boldsymbol{f_S(Z_A)}$" (cf. Eq. 10) cannot be different for different scale conversion methods. Thus, as already clarified (cf. Eq. 4d$^{///}$: ${}^{S}\boldsymbol{\varepsilon_i}{}^{(\text{Mtd-1})} = {}^{Z}\boldsymbol{\varepsilon_A}{}^{(\text{Mtd-1})}$), the estimates ${}^{S}\boldsymbol{r_i}{}^{(\text{Mtd-2})}$ and $\boldsymbol{z_A}{}^{(\text{Mtd-2})}$ should also be *equally* accurate: ${}^{S}\boldsymbol{\varepsilon_i}{}^{(\text{Mtd-2})} = {}^{Z}\boldsymbol{\varepsilon_A}{}^{(\text{Mtd-2})}$. In other words, even the findings here (i.e. which are based on the considerations of *ratio-of-ratios*) are in conformity with the previously pointed out fact[17,19] that the aid of any single ***Ai***-standard should cause the result (here: $\boldsymbol{z_A}$, or: $\boldsymbol{^{S}r_i}$) to be rather inaccurate: $^{Z}\boldsymbol{\varepsilon}_A{}^{(\text{Mtd-2})} > {}^{Z}\boldsymbol{\varepsilon}_A{}^{(\text{Mtd-1})}$; or: $^{S}\boldsymbol{\varepsilon}_i{}^{(\text{Mtd-2})} > {}^{S}\boldsymbol{\varepsilon}_i{}^{(\text{Mtd-1})}$.

•2. <u>**δ**-Scale conversion</u> ($\boldsymbol{Y_\delta} \xrightarrow{\boldsymbol{A1}} \boldsymbol{Z_\delta}$)

"$\boldsymbol{Z_\delta = f_\delta(Y_\delta, Y1_\delta)}$" should have description as:[17,19]

$$Z_\delta = \left(\frac{^{S}R_i}{^{D}R_i} - 1\right) = \left(\frac{^{S}R_i}{^{W}R_i} \times \frac{^{W}R_i}{^{A1}R_i} \times \frac{^{A1}R_i}{^{D}R_i} - 1\right) = \left(\frac{[Y_\delta + 1] \times [C1_\delta + 1]}{[Y1_\delta + 1]} - 1\right) \tag{11a}$$

And, therefore:

$$(z_\delta{}^{(\text{Mtd-2})} \pm {}^{Z}\varepsilon_\delta{}^{(\text{Mtd-2})}) = ([y_\delta \pm \varepsilon_\delta] \times [C1_\delta + 1])/(y1_\delta \pm \varepsilon1_\delta) \tag{11a$'$}$$

However, any "$\boldsymbol{Y_\delta = ([^{S}R_i/^{W}R_i] - 1)}$" type of relationship should a member of the F.2 family; i.e. (cf. Eqs. 4a$'$ and 7a): $\boldsymbol{\varepsilon_\delta} = \left(\left|\frac{^{S}\boldsymbol{R_i}}{^{S}\boldsymbol{R_i} - {}^{W}\boldsymbol{R_i}}\right| \times \boldsymbol{\varepsilon_A}\right) = \left(\left|\frac{^{S}\boldsymbol{R_i}}{^{S}\boldsymbol{R_i} - {}^{W}\boldsymbol{R_i}}\right| \times \boldsymbol{2} \times {}^{G}\boldsymbol{u}\right)$; and (one can verify that): $\boldsymbol{\varepsilon1_\delta} = \left(\left|\frac{^{A1}\boldsymbol{R_i}}{^{A1}\boldsymbol{R_i} - {}^{W}\boldsymbol{R_i}}\right| \times \boldsymbol{\varepsilon1_A}\right) = \left(\left|\frac{^{A1}\boldsymbol{R_i}}{^{A1}\boldsymbol{R_i} - {}^{W}\boldsymbol{R_i}}\right| \times \boldsymbol{2} \times {}^{G}\boldsymbol{u}\right)$.

Moreover, as already shown elsewhere[17], Eq. 11a should be characterized by the parameters (cf. Eq. 5) as: $\boldsymbol{M_{Y_\delta}^{Z_\delta} = \frac{^{S}R_i - {}^{W}R_i}{^{S}R_i - {}^{D}R_i}}$; and: $\boldsymbol{M_{Y1_\delta}^{Z_\delta} = -\frac{^{S}R_i(^{A1}R_i - {}^{W}R_i)}{^{A1}R_i(^{S}R_i - {}^{D}R_i)}}$; and thus (cf. Eq. 4, or so):[6]

$$^{Z}\varepsilon_\delta{}^{(\text{Mtd-2})} = \sum_{i=1}^{2} \left(\left|M_i^{Z_\delta}\right| \times \varepsilon_i\right) = \left[\left(\left|M_{Y_\delta}^{Z_\delta}\right| \times \boldsymbol{\varepsilon_\delta}\right) + \left(\left|M_{Y1_\delta}^{Z_\delta}\right| \times \boldsymbol{\varepsilon1_\delta}\right)\right] = \left[\left(\left|\frac{^{S}\boldsymbol{R_i}}{^{S}\boldsymbol{R_i} - {}^{D}\boldsymbol{R_i}}\right| \times \boldsymbol{\varepsilon_A}\right) + \left(\left|\frac{^{S}\boldsymbol{R_i}}{^{S}\boldsymbol{R_i} - {}^{D}\boldsymbol{R_i}}\right| \times \boldsymbol{\varepsilon1_A}\right)\right] = \left(\left|\frac{^{S}\boldsymbol{R_i}}{^{S}\boldsymbol{R_i} - {}^{D}\boldsymbol{R_i}}\right| \times [\boldsymbol{\varepsilon_A} + \boldsymbol{\varepsilon1_A}]\right) = \left(\left|\frac{^{S}\boldsymbol{R_i}}{^{S}\boldsymbol{R_i} - {}^{D}\boldsymbol{R_i}}\right| \times {}^{Z}\boldsymbol{\varepsilon_A}{}^{(\text{Mtd-2})}\right) \tag{4g}$$

Or, even for: $\boldsymbol{\varepsilon_A = \varepsilon1_A = 2\,^{G}u}$ (cf. further Eq. 4f$''$):

$$ {}^{Z}\varepsilon_{\delta}^{(\text{Mtd-2})} = \left(\left|\frac{{}^{S}R_i}{{}^{S}R_i - {}^{D}R_i}\right| \times [\varepsilon_A + \varepsilon 1_A]\right) = \left(\left|\frac{{}^{S}R_i}{{}^{S}R_i - {}^{W}R_i}\right| \times 4 \times {}^{G}u\right) = \left[\left(\left|\frac{{}^{S}R_i}{{}^{S}R_i - {}^{W}R_i}\right| \times {}^{Z}\varepsilon_A^{(\text{Mtd-2})}\right)\right. $$

$$ = (8.967 \times {}^{Z}\varepsilon_A^{(\text{Mtd-2})}) = (2 \times [8.967 \times 2 \times {}^{G}u]) = (2 \times {}^{Z}\varepsilon_{\delta}^{(\text{Mtd-1})}) \qquad (4g') $$

That is, "$z_{\delta}^{(\text{Mtd-2})}$" can turn out even twice more erroneous than "$z_{\delta}^{(\text{Mtd-1})}$".

•2.1. Conversion: $z_{\delta}^{(\text{Mtd-2})} \rightarrow {}^{S}r_i^{(\text{Mtd-2})}$

As "$z_{\delta} \rightarrow {}^{S}r_i$" conversion formula cannot vary for varying the scale conversion method, it could here again be shown that (cf. **Mtd-1, viz.** Eq. 4d/// or Eq. 7b or so): ${}^{S}\varepsilon_i^{(\text{Mtd-2})} = {}^{Z}\varepsilon_A^{(\text{Mtd-2})} = ({}^{Z}\varepsilon_{\delta}^{(\text{Mtd-2})}/[\,|({}^{S}R_i - {}^{D}R_i)/{}^{S}R_i|\,]) = ({}^{Z}\varepsilon_{\delta}^{(\text{Mtd-2})}/8.9671)$.

3.3.2.2 **Mtd-3:** aid of two different ***Ai***-standards (***A*1**, ***A*2**)

■1. Scale conversion of ratio-of-ratios ($Y_A \xrightarrow{A1,A2} Z_A$)

Let: $Y2_A = ({}^{A2}R_i/{}^{W}R_i)$. Then the relation "$Z_A = f_A(Y_A, Y1_A, Y2_A)$" can, like Eq. 9a, be derived as follows:

$$ Z_A = \left(\frac{{}^{S}R_i}{{}^{D}R_i}\right) = \left(\frac{{}^{S}R_i}{{}^{W}R_i} \times \frac{{}^{W}R_i}{{}^{A1}R_i - {}^{A2}R_i} \times \frac{{}^{A1}R_i - {}^{A2}R_i}{{}^{D}R_i}\right) $$

$$ = \left(Y_A \times \frac{C1_A - C2_A}{Y1_A - Y2_A}\right) = \left(Y_A \times \frac{C1_{\delta} - C2_{\delta}}{Y1_A - Y2_A}\right) \qquad (9b) $$

That is, in terms of estimates:

$$ (z_A^{(Mtd-3)} \pm {}^{Z}\varepsilon_A^{(Mtd-3)}) = \left([y_A \pm \varepsilon_A] \times \frac{C1_{\delta} - C2_{\delta}}{[y1_A \pm \varepsilon 1_A] - [y2_A \pm \varepsilon 2_A]}\right) \qquad (9b') $$

where:: $C2_{\delta} = [({}^{A2}R_i/{}^{D}R_i) - 1] = [{}^{A2/D}C1_A - 1]$); i.e. $C2_{\delta}$ is the "***A*2** vs. ***D***" calibration constant; ${}^{Z}\varepsilon_A^{(\text{Mtd-3})}$ represents the uncertainty of (the present *method-specific* estimate) $z_A^{(\text{Mtd-3})}$; and $\varepsilon 2_A$ stands (like $\varepsilon_A$ and $\varepsilon 1_A$) for the *lab-established* uncertainty of the absolute estimate "$y2_A$".

Further, Eq. 9b could be seen to be characterized by the parameters (cf. Eq. 5) as:

$\boldsymbol{M^{Z_A}_{Y_A}}$ = **1**; but**:** $\boldsymbol{M^{Z_A}_{Y1_A}}$ = –($^{A1}\boldsymbol{R_i}$/[$^{A1}\boldsymbol{R_i}$ – $^{A2}\boldsymbol{R_i}$]) and**:** $\boldsymbol{M^{Z_A}_{Y2_A}}$ = ($^{A2}\boldsymbol{R_i}$/[$^{A1}\boldsymbol{R_i}$ – $^{A2}\boldsymbol{R_i}$]).

That is (even though both Eq. 9 and Eq. 9a belong to the F.1 model family) Eq. 9b is a member of the F.2 family. Therefore, the choice of even (the present ratio-of-ratios' based formula) Eq. 9b as the IRMS evaluation model *should be* ***risky***. Clearly, the reason is that the uncertainty "$^{Z}\boldsymbol{\varepsilon_A}^{\textbf{(Mtd-3)}}$" should not only be higher (than either "$^{Z}\boldsymbol{\varepsilon_A}^{\textbf{(Mtd-1)}}$" or "$^{Z}\boldsymbol{\varepsilon_A}^{\textbf{(Mtd-2)}}$") but also be dependent on the ***Ai***-standards to be used (cf. Eq. 4 or so):

$$^{Z}\boldsymbol{\varepsilon_A}^{\text{(Mtd-3)}} = [(|M^{Z_A}_{Y_A}| \times \varepsilon_A) + (|M^{Z_A}_{Y1_A}| \times \varepsilon1_A) + (|M^{Z_A}_{Y2_A}| \times \varepsilon2_A)]$$

$$= [\varepsilon_A + (\left|\frac{^{A1}R_i}{^{A1}R_i - {}^{A2}R_i}\right| \times \varepsilon1_A) + (\left|\frac{^{A2}R_i}{^{A1}R_i - {}^{A2}R_i}\right| \times \varepsilon2_A)] \qquad \text{(4h)}$$

Or (for [cf. e.g. Eq. 4a$^{/}$)**:** $\boldsymbol{\varepsilon_A = \varepsilon1_A = \varepsilon2_A = 2^{G}u}$):

$$^{Z}\boldsymbol{\varepsilon_A}^{\text{(Mtd-3)}} = ([2 \times \left(1+\left|\frac{^{A1}R_i + {}^{A2}R_i}{^{A1}R_i - {}^{A2}R_i}\right|\right)] \times {}^{G}u) = (^{Z}[\boldsymbol{UF}]_{\boldsymbol{A}}^{\textbf{(Mtd-3)}} \times {}^{G}u) \qquad \text{(4h}^{/}\text{)}$$

Clearly, $^{Z}[\boldsymbol{UF}]_{\boldsymbol{A}}^{\textbf{(Mtd-3)}}$ should be **>4** (and hence cause**:** $^{Z}\boldsymbol{\varepsilon_A}^{\textbf{(Mtd-3)}}$ > $^{Z}\boldsymbol{\varepsilon_A}^{\textbf{(Mtd-2)}}$); i.e. "$\boldsymbol{z_A}^{\textbf{(Mtd-3)}}$" should be more erroneous than even "$\boldsymbol{z_A}^{\textbf{(Mtd-2)}}$".

■ ($\boldsymbol{z_A}^{\text{(Mtd-3)}}$ → $^{S}\boldsymbol{r_i}^{\text{(Mtd-3)}}$) Conversion

The formula "$^{S}\boldsymbol{R_i}$ = $\boldsymbol{f_S}(\boldsymbol{Z_A})$" is fixed (cf. Eq. 10). Thus, like other methods (cf. e.g. Eq. 4d), it could be shown that**:** $^{S}\boldsymbol{\varepsilon_i}^{\textbf{(Mtd-3)}}$ = $^{Z}\boldsymbol{\varepsilon_A}^{\textbf{(Mtd-3)}}$.

■ <u>**δ**-Scale conversion </u>($\boldsymbol{Y_\delta} \xrightarrow{\boldsymbol{A1,A2}} \boldsymbol{Z_\delta}$)

"$\boldsymbol{Z_\delta} = \boldsymbol{f_\delta}(\boldsymbol{Y_\delta}, \boldsymbol{Y1_\delta}, \boldsymbol{Y2_\delta})$" should, like Eq. 9b, take the description as:

$$Z_\delta = \left(\frac{^{S}R_i}{^{D}R_i} - 1\right) = \left(\left[\frac{^{S}R_i}{^{W}R_i} \times \frac{^{W}R_i}{^{A1}R_i - {}^{A2}R_i} \times \frac{^{A1}R_i - {}^{A2}R_i}{^{D}R_i}\right] - 1\right)$$

$$= \left( \left[ (Y_\delta + 1) \times \frac{(C1_\delta - C2_\delta)}{(Y1_\delta - Y2_\delta)} \right] - 1 \right) \tag{11b}$$

Therefore:

$$(z_\delta{}^{(Mtd-3)} \pm {}^{Z}\varepsilon_\delta{}^{(Mtd-3)}) = \left( [y_\delta \pm \varepsilon_\delta] \times \frac{[C1_\delta - C2_\delta]}{[y1_\delta \pm \varepsilon 1_\delta] - [y2_\delta \pm \varepsilon 2_\delta]} \right) \tag{11b$'$}$$

However, it should be pointed out that, although the literature[14,15] based "$\boldsymbol{Z_\delta = f_\delta(Y_\delta, Y1_\delta)}$" formula is identical with Eq. 11a, the commonly used "$\boldsymbol{Z_\delta = f_\delta(Y_\delta, Y1_\delta, Y2_\delta)}$" formula is as follows:[14,15]

$$Z_\delta = \left( \left[ (Y_\delta - Y2_\delta) \times \frac{(C1_\delta - C2_\delta)}{(Y1_\delta - Y2_\delta)} \right] + C2_\delta \right) \tag{12}$$

However, it is already clarified elsewhere[17,19] that, as "$\boldsymbol{C1_\delta}$ and $\boldsymbol{C2_\delta}$" are *constants* and "$\boldsymbol{Z1_\delta}$ and $\boldsymbol{Z2_\delta}$" are *variables* (and thus, as**:** $[(\boldsymbol{{}^{A1}R_i/{}^{D}R_i}) \times (\boldsymbol{{}^{A2}R_i/{}^{W}R_i})] \neq [(\boldsymbol{{}^{A1}R_i/{}^{W}R_i}) \times (\boldsymbol{{}^{A2}R_i/{}^{D}R_i})]$); the right side of Eq. 12 cannot be reduced to the left side "$(\boldsymbol{[{}^{S}R_i/{}^{D}R_i] - 1})$", i.e. Eq. 12 cannot represent any possible form of the "$(\boldsymbol{Y_\delta \xrightarrow{A1,A2} Z_\delta})$" conversion process.

Anyway, Eq. 11b should, like any simple **δ**-expression (e.g. Eq. 2), belong to the F.2 model family. Thus, recollecting that (cf. Eqs. 4a$'$ and 7a)**:** $\boldsymbol{\varepsilon_\delta} = (|(\boldsymbol{{}^{S}R_i/[{}^{S}R_i - {}^{W}R_i]})| \times \boldsymbol{\varepsilon_A})$, $\boldsymbol{\varepsilon 1_\delta} = (|(\boldsymbol{{}^{A1}R_i/[{}^{A1}R_i - {}^{W}R_i]})| \times \boldsymbol{\varepsilon 1_A})$ and**:** $\boldsymbol{\varepsilon 2_\delta} = (|(\boldsymbol{{}^{A2}R_i/[{}^{A2}R_i - {}^{W}R_i]})| \times \boldsymbol{\varepsilon 2_A})$; the uncertainty of evaluation $\boldsymbol{{}^{Z}\varepsilon_\delta{}^{(Mtd\text{-}3)}}$ could be shown to be decided as (cf. Eq. 4 and also Eq. 4h):[17,19]

$${}^{Z}\varepsilon_\delta{}^{(Mtd\text{-}3)} = ([|M_{Y_\delta}^{Z_\delta}| \times \varepsilon_\delta] + [|M_{Y1_\delta}^{Z_\delta}| \times \varepsilon 1_\delta] + [|M_{Y2_\delta}^{Z_\delta}| \times \varepsilon 2_\delta])$$

$$= ([\left|\frac{{}^{S}R_i - {}^{W}R_i}{{}^{S}R_i - {}^{D}R_i}\right| \times \varepsilon_\delta] + [\left|\frac{-{}^{S}R_i \times ({}^{A1}R_i - {}^{W}R_i)}{({}^{A1}R_i - {}^{A2}R_i) \times ({}^{S}R_i - {}^{D}R_i)}\right| \times \varepsilon 1_\delta] + [\left|\frac{-{}^{S}R_i \times ({}^{A2}R_i - {}^{W}R_i)}{({}^{A1}R_i - {}^{A2}R_i) \times ({}^{S}R_i - {}^{D}R_i)}\right| \times \varepsilon 2_\delta])$$

$$= (\left|\frac{{}^{S}R_i}{{}^{S}R_i - {}^{D}R_i}\right| \times \boldsymbol{[\varepsilon_A} + (\left|\frac{{}^{A1}R_i}{{}^{A1}R_i - {}^{A2}R_i}\right| \times \boldsymbol{\varepsilon 1_A}) + (\left|\frac{{}^{A2}R_i}{{}^{A1}R_i - {}^{A2}R_i}\right| \times \boldsymbol{\varepsilon 2_A})\boldsymbol{]})$$

$$= \left( \left| \frac{{}^{S}R_i}{{}^{S}R_i - {}^{D}R_i} \right| \times {}^{Z}\boldsymbol{\varepsilon}_A{}^{\mathbf{(Mtd\text{-}3)}} \right) = \left( \left| \frac{{}^{S}R_i}{{}^{S}R_i - {}^{D}R_i} \right| \times {}^{S}\boldsymbol{\varepsilon}_i{}^{\mathbf{(Mtd\text{-}3)}} \right) = \left( 8.9671 \times {}^{S}\boldsymbol{\varepsilon}_i{}^{\mathbf{(Mtd\text{-}3)}} \right) \quad (4i)$$

Or, even for**:** $\boldsymbol{\varepsilon}_A = \boldsymbol{\varepsilon}\mathbf{1}_A = \boldsymbol{\varepsilon}\mathbf{2}_A = \mathbf{2}\,{}^{G}\boldsymbol{u}$ (see also Eq. 4h/):

$${}^{Z}\varepsilon_\delta{}^{(Mtd\text{-}3)} = \left( \left| \frac{{}^{S}R_i}{{}^{S}R_i - {}^{D}R_i} \right| \times \mathbf{2} \times {}^{G}\boldsymbol{u} \times \mathbf{[1} + \left| \frac{{}^{A1}R_i}{{}^{A1}R_i - {}^{A2}R_i} \right| + \left| \frac{{}^{A2}R_i}{{}^{A1}R_i - {}^{A2}R_i} \right| \mathbf{]} \right)$$

$$= \left( \left| \frac{{}^{S}R_i}{{}^{S}R_i - {}^{D}R_i} \right| \times \mathbf{2} \times \mathbf{[1} + \left| \frac{{}^{A1}R_i + {}^{A2}R_i}{{}^{A1}R_i - {}^{A2}R_i} \right| \mathbf{]} \right) \times {}^{G}\boldsymbol{u}$$

$$= \left( \left[ \left| \frac{{}^{S}R_i}{{}^{S}R_i - {}^{D}R_i} \right| \times {}^{Z}[\boldsymbol{UF}]_A{}^{\mathbf{(Mtd\text{-}3)}} \right] \times {}^{G}\boldsymbol{u} \right) = \left( {}^{Z}[\boldsymbol{UF}]_\delta{}^{\mathbf{(Mtd\text{-}3)}} \times {}^{G}\boldsymbol{u} \right) \quad (4i^{/})$$

It is clarified above that "${}^{Z}[\boldsymbol{UF}]_A{}^{\mathbf{(Mtd\text{-}3)}}$" should itself be >4. Therefore, for present known case, "${}^{Z}[\boldsymbol{UF}]_\delta{}^{\mathbf{(Mtd\text{-}3)}}$" should be as high as >36; i.e.**:** ${}^{Z}\boldsymbol{\varepsilon}_\delta{}^{\mathbf{(Mtd\text{-}3)}} > \mathbf{(36 \times {}^{G}\boldsymbol{u})}$.

Moreover, it is already exemplified elsewhere[17] that *any method specific* error-ratio as "${}^{Z}\mathbf{Đ}_\delta/{}^{S}\mathbf{Đ}_i$" does turn out to be a constant equaling to "$|\,({}^{S}\boldsymbol{R}_i/[{}^{S}\boldsymbol{R}_i - {}^{D}\boldsymbol{R}_i])\,|$".

However, the findings**:** (i) ${}^{Z}\boldsymbol{\varepsilon}_A{}^{\mathbf{(Mtd\text{-}1)}} < {}^{Z}\boldsymbol{\varepsilon}_A{}^{\mathbf{(Mtd\text{-}2)}} < {}^{Z}\boldsymbol{\varepsilon}_A{}^{\mathbf{(Mtd\text{-}3)}}$ (cf. the different *ratio-of-ratios* based scale conversion methods as Eqs. 9, 9a and 9b); (ii) ${}^{Z}\boldsymbol{\varepsilon}_\delta{}^{\mathbf{(Mtd\text{-}1)}} < {}^{Z}\boldsymbol{\varepsilon}_\delta{}^{\mathbf{(Mtd\text{-}2)}} < {}^{Z}\boldsymbol{\varepsilon}_\delta{}^{\mathbf{(Mtd\text{-}3)}}$ (cf. the **δ**-scale conversion methods as Eq. 11, 11a and 11b); and**:** (iii) ${}^{S}\boldsymbol{\varepsilon}_i{}^{\mathbf{(Mtd\text{-}1)}} < {}^{S}\boldsymbol{\varepsilon}_i{}^{\mathbf{(Mtd\text{-}2)}} < {}^{S}\boldsymbol{\varepsilon}_i{}^{\mathbf{(Mtd\text{-}3)}}$ (i.e. in terms of even sample isotopic ratio obtained by either Eq. 10 or 10a); should represent the simple fact that the uncertainty of any evaluation should be proportional to the number of individual measurements to be involved. In other words, the claim[11,14,15,18] that "${}^{Z}\boldsymbol{\varepsilon}_\delta{}^{\mathbf{(Mtd\text{-}2)}} > {}^{Z}\boldsymbol{\varepsilon}_\delta{}^{\mathbf{(Mtd\text{-}3)}}$" can never represent any real world event.

What is however significant is that, ***irrespective*** of scale conversion method, the *absolute* estimate (either "$\boldsymbol{z}_A$" or "${}^{S}\boldsymbol{r}_i$") is shown to be **more** accurate than the **δ**-*estimate* "$\boldsymbol{z}_\delta$". Furthermore, the *ratio* of "**δ**-to-**absolute**" *errors* (i.e. ratio of achievable "δ-to-absolute" accuracies and/ or uncertainty factors) should, *for using any scale conversion method*, be a ***constant*** *as* "$|(^{S}\boldsymbol{R_i}/[^{S}\boldsymbol{R_i} - {}^{D}\boldsymbol{R_i}])|$; cf. Eq. 7b". That is the said constant (error-ratio) should be *prefixed by* ***only*** the sample "***S***" and the reference-standard "***D***" involved. Therefore, if "$^{S}\boldsymbol{R_i}$" should happen to be very close to "$^{D}\boldsymbol{R_i}$", then the evaluated *absolute* estimate ($\boldsymbol{z_A}$, or even $^{S}\boldsymbol{r_i}$) should be highly **accurate**, but the corresponding **δ**-*estimate* "$\boldsymbol{z_\delta}$" might turn out to be highly *erroneous*.

## CONCLUSIONS

The above study clarifies that even simply *specifying* a mathematical relationship "$\boldsymbol{Y_d} = \boldsymbol{f_d}(\{\boldsymbol{X_i}\})$" should mean[3] *evaluating* the proposed function "$\boldsymbol{f_d}$" as a model for any purpose whatever. The reason is, as clarified above, that the "$\boldsymbol{f_d}$" specific *individual* and, hence the *collective*, input-to-output *variation-**rates*** ($\boldsymbol{M_i^d}$ and $[\boldsymbol{UF}]_{\boldsymbol{d}}$, respectively) should also thus get specified: $\boldsymbol{M_i^d} = \left(\frac{\partial Y_d}{\partial X_i}\right)\left(\frac{X_i}{Y_d}\right)$, with: $\boldsymbol{i} = 1, 2 \ldots \boldsymbol{N}$; and: $[\boldsymbol{UF}]_{\boldsymbol{d}} = \sum_{i=1}^{N}\left|\boldsymbol{M_i^d}\right|$.

In other words, it is signified above that any conceivable *input-output **variation*** should, in nature, be purely systematic[3] (i.e. "$\boldsymbol{f_d}$" specific). It is thus elaborated that, for a given variation (viz.) "$^{G}\boldsymbol{u} = 1\%$" in the measurable and/ or input variables ($\boldsymbol{X_i}$, with: $\boldsymbol{i} = 1, 2 \ldots \boldsymbol{N}$), the possible *maximum value* ($\varepsilon_{\boldsymbol{d}}$) of output-variation can 'a priori' be ascertained as: $\varepsilon_{\boldsymbol{d}} = ([\boldsymbol{UF}]_{\boldsymbol{d}} \times {}^{G}\boldsymbol{u}) = [\boldsymbol{UF}]_{\boldsymbol{d}}\boldsymbol{\%}$. Clearly, the *smaller* should be the "$[\boldsymbol{UF}]_{\boldsymbol{d}}$" value (viz.: $[\boldsymbol{UF}]_{\boldsymbol{d}} << 1$) the ***better*** suitable the model "$\boldsymbol{f_d}$" be for any purpose whatsoever.

Depending upon the exact description of "$\boldsymbol{f_d}$", one or the other "$\boldsymbol{M_i^d}$", and hence "$[\boldsymbol{UF}]_{\boldsymbol{d}}$", might turn out to be decided by the system-specific $\boldsymbol{X_i}$-value(s). However, the above study supports the classification[3] of all conceivable models into two families: (i) all models but for each of which cases "$[\boldsymbol{UF}]_{\boldsymbol{d}}$" should be ***independent*** *of* "*system* (viz. $\boldsymbol{X_i}$)" be a single family (**F.1**) members; and**:** (ii) all others (i.e. for each of which cases, "$[\boldsymbol{UF}]_{\boldsymbol{d}}$" should ***vary*** as a function of $\boldsymbol{X_i(s)}$) or so should belong to another single family (**F.2**) only.

The significance of the model-families is that, while the **F.1** *modeling/ output accuracy* (and hence the accuracy of driving any corresponding model based machine) should *solely* be decided by achievable $\boldsymbol{X_i}$-measurement accuracies **(**$\varepsilon_{\boldsymbol{d}}^{\mathbf{(F.1)}} = \sum_{i=1}^{N} \left(\left|\boldsymbol{M_i^d}\right| \times \boldsymbol{u_i}\right) = \sum_{i=1}^{N} \boldsymbol{u_i}$; and/ or**:** $\varepsilon_{\boldsymbol{d}}^{\mathbf{(F.1)}} = [\boldsymbol{N} \times {}^{G}\boldsymbol{u}]$, with**:** $\boldsymbol{u_1} = \boldsymbol{u_2} \ldots = \boldsymbol{u_N} = {}^{G}\boldsymbol{u}$**)**; the achievable **F.2** modeling accuracy is shown to be governed as**:** $\varepsilon_{\boldsymbol{d}}^{\mathbf{(F.2)}} = \boldsymbol{f_d}(\{\boldsymbol{X_i},\boldsymbol{u_i}\})$.

Clearly, any F.2 model should generally be risky to employ in practice. This is because that, for *specific* $\boldsymbol{X_i}$-value(s), the possible (net) $\boldsymbol{X_i}$-measurement-error might happen to even be *reduced* in the process of defining the output-error (and thus may offer *an unexpected control* over a corresponding output based machine). However, for certain **other** possible value(s) of system defining $\boldsymbol{X_i(s)}$, the net input-error might even be **enhanced** as the output-error (i.e. may lead a corresponding machine to disastrous consequences).

A simple example of the F.1 models is the elementary mathematical ***expression*** of any *absolute* ratio **(**viz.**:** $\boldsymbol{Y_A} = [{}^{S}\boldsymbol{R_i}/{}^{W}\boldsymbol{R_i}]$, with "${}^{S}\boldsymbol{R_i}$ and ${}^{W}\boldsymbol{R_i}$" representing the well-known IRMS sample ***S*** and working-lab-reference ***W*** specific $\boldsymbol{i^{th}}$ isotopic abundance-ratios, respectively**)**; and that of the F.2 models is (any *differential* "**δ**" ratio)**:** $\boldsymbol{Y_\delta} = ([{}^{S}\boldsymbol{R_i}/[{}^{W}\boldsymbol{R_i}] - \mathbf{1}) = (\boldsymbol{Y_A} - \mathbf{1})$. However, what is important is that, for any specific case of ***S*** and ***W*** measurements, the ***ratio*** of the possible *true-errors* $\boldsymbol{Đ_A}$ and $\boldsymbol{Đ_\delta}$ **(**and hence of the uncertainties, $\boldsymbol{\varepsilon_A}$ and $\boldsymbol{\varepsilon_\delta}$, of the evaluated unknown *absolute* and **δ**-*ratios*, $\boldsymbol{y_A}$ and $\boldsymbol{y_\delta}$, respectively**)** should be ***prefixed*** as:

$$\left({}^{Đ_\delta}/_{Đ_A}\right) = \left({}^{\varepsilon_\delta}/_{\varepsilon_A}\right) = \left({}^{[UF]_\delta}/_{[UF]_A}\right) = \left|\frac{{}^{S}R_i}{{}^{S}R_i - {}^{W}R_i}\right|$$

Further, "$|{}^{S}\boldsymbol{R_i} - {}^{W}\boldsymbol{R_i}| \rightarrow \mathbf{0}$" is a requirement for IRMS *measurement*. Therefore, any **δ**-estimate ($\boldsymbol{y_\delta}$) should clearly be ***more*** *erroneous* than the corresponding absolute estimate ($\boldsymbol{y_A}$) and, at least for a case of very close isotopic compositions (ICs) of ***S*** and ***W***, turn out *highly* erroneous.

Again, the **δ**-*scale conversion*: $\boldsymbol{y_\delta} \rightarrow \boldsymbol{z_\delta}$ (with: $\boldsymbol{z_\delta}$ = [$({}^{S}\boldsymbol{r_i}/{}^{D}\boldsymbol{R_i}) - \mathbf{1}$] = [$\boldsymbol{z_A} - \mathbf{1}$]; and "***D***" as the recommended reference-standard**)** is unavoidable for *reporting* any IRMS result (**δ**-estimate). However, the scale conversion of *ratio-of-ratios* "$\boldsymbol{y_A} \rightarrow \boldsymbol{z_A}$" is shown above to rather be a simpler task. Further, "${}^{D}\boldsymbol{R_i}$" should be known. Therefore, the *sample isotopic abundance ratio* (${}^{S}\boldsymbol{r_i}$) can also be evaluated, i.e. from "$\boldsymbol{z_\delta}$" as: $({}^{S}\boldsymbol{r_i} \pm {}^{S}\varepsilon_i) = ({}^{D}\boldsymbol{R_i} \times [(\boldsymbol{z_\delta} \pm {}^{Z}\varepsilon_\delta) + 1])$; and/ or from "$\boldsymbol{z_A}$" as: $({}^{S}\boldsymbol{r_i} \pm {}^{S}\varepsilon_i) = ({}^{D}\boldsymbol{R_i} \times [\boldsymbol{z_A} \pm {}^{Z}\varepsilon_A])$.

However, the above study supplements the previously reported fact[17,19] that the scale conversion with the aid of any single auxiliary (***Ai***) standard should, though help avoid the requirement of calibrating "***W***", cause the desired result (viz. **δ**-estimate "$\boldsymbol{z_\delta}$", or even any *absolute* estimate "$\boldsymbol{z_A}$" or "${}^{S}\boldsymbol{r_i}$") to be more erroneous than that to be obtained by the ***W***-calibration method. In other words, the claim[11, 14,15,18] that the process "($\boldsymbol{Y_\delta} \xrightarrow{\boldsymbol{A1,A2}} \boldsymbol{Z_\delta}$)", rather than "($\boldsymbol{Y_\delta} \xrightarrow{\boldsymbol{A1}} \boldsymbol{Z_\delta}$)", should ensure "$\boldsymbol{z_\delta}$" to be accurate is simply baseless. The employing of increasing number of ***Ai***-standards should require increasing number of measurements, and hence subject any scale converted data (either $\boldsymbol{z_\delta}$ or $\boldsymbol{z_A}$) to increasing number of different measurement-errors.

Moreover, the present study emphasizes the fact[17] that, for whatsoever *method* might be chosen, the ratio of "**δ**-to-**absolute**" scale conversion uncertainties (i.e. "${}^{Z}\varepsilon_\delta/{}^{Z}\varepsilon_A$", or "${}^{Z}\varepsilon_\delta/{}^{S}\varepsilon_i$") and/ or the ratio of possible true-errors (i.e.: "${}^{Z}Đ_\delta/{}^{Z}Đ_A$", or "${}^{Z}Đ_\delta/{}^{S}\varepsilon_i$") should be ***prefixed*** by, only, the sample ***S*** and the standard ***D*** involved:

$$\frac{{}^{Z}Đ_{\delta}}{{}^{Z}Đ_{A}} = \frac{{}^{Z}Đ_{\delta}}{Đ_{A}} = \frac{{}^{Z}Đ_{\delta}}{{}^{S}Đ_{i}} = \frac{{}^{Z}\varepsilon_{\delta}}{{}^{Z}\varepsilon_{A}} = \frac{{}^{Z}\varepsilon_{\delta}}{\varepsilon_{A}} = \frac{{}^{Z}\varepsilon_{\delta}}{{}^{S}\varepsilon_{i}} = \left| \frac{{}^{S}R_{i}}{{}^{S}R_{i} - {}^{D}R_{i}} \right|$$

Again, "$^{D}\boldsymbol{R_i} \approx {}^{S}\boldsymbol{R_i} \approx {}^{W}\boldsymbol{R_i}$." is a basic requirement for any IRMS evaluation. Thus, the **δ**-estimate $\boldsymbol{z_\delta}$ might even turn out, i.e. for a possible case of very similar ICs of ***S*** and ***D***, *misleading*. This further emphasizes the fact[17,19] that different possible *absolute* lab-results (either ratio-of-ratios**:** $\boldsymbol{z_A}^{\textbf{(Lab-1)}}$, $\boldsymbol{z_A}^{\textbf{(Lab-2)}}$ **…**; or sample isotopic ratios**:** $^{S}\boldsymbol{r_i}^{\textbf{(Lab-1)}}$, $^{S}\boldsymbol{r_i}^{\textbf{(Lab-2)}}$ **…**) should more accurately represent the sample "***S***" and/ or more closely be intercomparable than the corresponding **δ**-estimates ($\boldsymbol{z_\delta}^{\textbf{(Lab-1)}}$, $\boldsymbol{z_\delta}^{\textbf{(Lab-2)}}$ **…**).

Essentially, the basic IRMS evaluation principle "$\boldsymbol{Z_\delta = f_\delta(Y_\delta)}$" is shown to be worth reformulating as either "$^{S}\boldsymbol{R_i = f_S(Z_\delta) = f_S(f_\delta(Y_\delta))}$" or simply as "$\boldsymbol{Z_A = f_A(Y_A)}$".

Of course, it should generally be difficult to calibrate any possible lab-specific working-reference (***W***). Therefore, for scale conversion, the aid of a suitable ***Ai***-standard (and hence the allowance of certain *additional* uncertainty in the result) should also be unavoidable. Yet, as shown above, the replacement of the **δ**-ratio based evaluation process "$\boldsymbol{Y_\delta \xrightarrow{A1} Z_\delta}$" by the absolute-ratio based evaluation process as either **"**$\boldsymbol{Y_\delta \xrightarrow{A1} Z_\delta \rightarrow {}^{S}R_i}$**"** or simply "$\boldsymbol{Y_A \xrightarrow{A1} Z_A}$" should help avoid the reporting of (relatively or even highly) erroneous **δ**-results. Moreover, either "$\boldsymbol{Y_A \rightarrow Z_A}$" or "$\boldsymbol{Y_A \xrightarrow{A1} Z_A}$" should mean simplification of required computations.

## REFERENCES

1. McKinney, C. R., McCrea, J. M., Epstein, S, Allen H. A. & Urey U. C. *Rev. Sci. Insttum*. **21**, 724 (1950).
2. IAEA-TECDOC-825, IAEA, Vienna (1995).
3. Datta, B. P. *arXiv***:0712.1732**.
4. ISO, *Guide to the Expression of Uncertainty in Measurement* (1995).
5. Scarborough J. B. *Numerical Mathematical Analysis*, Oxford & IBH Publishing Co., Kolkata (1966).
6. Datta, B. P. *arXiv***:0909.1651**.
7. IAEA-TECDOC-1247, IAEA, Vienna (2001).
8. Kerstel, E. R. Th., Trigt, R. van, Dam, N., Reuss, J. & Meijer, H. A. *Anal. Chem.* **71**, 5297 (1999).
9. Kerstel, E. R. Th. & Gianfrani, L. *Appl. Physics B* **92**, 439 (2008).
10. Coplen T. B. *Pure and Appl. Chem.* **66**, 273 (1994).
11. Coplen T. B., Brand, W. A., Gehre, M., Groning, M., Meijer, H.A., Toman, B. & Verkouteren, R. M. *Anal. Chem.* **78**, 2439 (2006).
12. Verkouteren R. M., Klouda G. A. & Currie L. A. *IAEA-TECDOC*-825, IAEA, Vienna, **1995**, pp. 111-129.
13. Allison, C. E., Francey, R. J. & Meijer H. A. J. *IAEA-TECDOC*-825, p. 155, IAEA, Vienna (1995).
14. Verkouteren, R. M. & Lee, J. N. *Fresenius J. Anal. Chem.* **370**, 803 (2001).
15. Paul, D., Skrzypek, G. & Forizs, I. *Rapid Commun. Mass Spectrom.* **21**, 3006 (2007).

16. Gonfiantini, R., Stichler, W. & Rozanski, K. *IAEA-TECDOC*-825, p. 13, IAEA, Vienna (1995).

17. Datta, B. P. *arXiv***:1401.1094**.

18. Gonfiantini R. *Nature* **271**, 534 (1978).

19. Datta, B. P. *arXiv***:1101.0973**.

20. Groning, M., Frohlich, K., Regge, P. De & Danesi P. P. INTENDED USE of THE IAEA REFERENCE MATERIALS PART II, p. 5, IAEA, Vienna (2009).

21. Vienna (2009).

**Table 1.** Possible *measured* estimates ($^{S}r_i$ and $^{W}r_i$, with errors $\Delta_S$ and $\Delta_W$) of the $^{2}H/^{1}H$ isotopic standards "IAEA-CH-7 and GISP" (respectively) and the corresponding *modeled*, i.e. "***S/W***" *absolute* and *differential*, estimates ($y_A$ and $y_\delta$; with errors $\text{Đ}_A$ and $\text{Đ}_\delta$, respectively)

| LAB No | Example No. | Measured "***S*** or ***W***" specific isotopic abundance ratio ($r_i$) (Measurement error Δ) | | LAB reflected uncertainty "$u_i$" (%) | Output estimate "$y_d$" (output-error $\text{Đ}_d$) | | Observed error-ratio $\frac{\lvert \text{Đ}_\delta \rvert}{\lvert \text{Đ}_A \rvert}$ | Predicted $\frac{\lvert \text{Đ}_\delta \rvert}{\lvert \text{Đ}_A \rvert}$ |
|---|---|---|---|---|---|---|---|---|
| | | $^{S}r_i \times 10^5$ ($\Delta_S \times 10^2$) | $^{W}r_i \times 10^5$ ($\Delta_W \times 10^2$) | | Absolute ratio ($y_A$) ($\text{Đ}_A \times 10^2$) | $^{S/W}\delta$-ratio ($y_\delta$) ($\text{Đ}_\delta \times 10^2$) | | |
| 1 | 0 | 14.013260 (0) | 12.62076552 (0) | - | 1.1103336 (0) | 0.1103336 (0) | 0 | 0 |
| | 1 | 14.01606265 (0.02) | 12.62215380 (0.011) | 0.02 | 1.1104335 (0.009) | 0.1104335 (0.0906) | 10.0634 | 10.0634 |
| | 2 | 14.01157841 (–0.012) | 12.62291105 (0.017) | | 1.1100117 (–0.029) | 0.1100117 (–0.292) | 10.0634 | |
| | 3 | 14.01424093 (0.007) | 12.62240622 (0.013) | | 1.1102670 (–0.006) | 0.1102670 (–0.0604) | 10.0634 | |
| | 4 | 14.01550212 (0.016) | 12.61849378 (–0.018) | | 1.1107112 (0.034) | 0.1107112 (0.3422) | 10.0634 | |
| | 5 | 14.01059748 (–0.019) | 12.62013448 (–0.005) | | 1.1101781 (–0.014) | 0.1101781 (–0.1409) | 10.0634 | |
| 2 | 0 | 14.013260 (0) | 12.62076552 (0) | | 1.1103336 (0) | 0.1103336 (0) | 0 | 0 |
| | 1 | 14.00345072 (–0.07) | 12.61079512 (–0.079) | 0.12 | 1.1104336 (0.009) | 0.1104336 (0.0906) | 10.0634 | 10.0634 |
| | 2 | 13.99896647 (–0.102) | 12.61142615 (–0.074) | | 1.1100225 (–.02802) | 0.1100225 (–0.282) | 10.0634 | |
| | 3 | 14.00106846 (–0.087) | 12.61066891 (–0.08) | | 1.1102558 (–0.007) | 0.1102558 (–0.0705) | 10.0634 | |
| | 4 | 14.00289019 (–0.074) | 12.80688268 (–0.11) | | 1.1107338 (0.03604) | 0.1107338 (0.3627) | 10.0634 | |
| | 5 | 13.99644409 (–0.12) | 12.60814475 (–0.10) | | 1.1101113 (–.02002) | 0.1101113 (–0.2015) | 10.0634 | |

**Table 2.** "Mtd-1" specific *scale converted* results $\boldsymbol{z_A}$ and $\boldsymbol{z_\delta}$ (i.e. corresponding to the possible *lab*-estimates $\boldsymbol{y_A}$ and $\boldsymbol{y_\delta}$, respectively, in Table 1); and also the sample isotopic ratio ($^{S}\boldsymbol{r_i}$) obtained by the process as either "$\boldsymbol{Z_A} \rightarrow {}^{S}\boldsymbol{R_i}$ (cf. Eq. 10)" or "$\boldsymbol{Z_\delta} \rightarrow {}^{S}\boldsymbol{R_i}$ (cf. Eq. 10a)"

| Lab No. | Example No. | Ratio-of-ratios | | Differential (**δ**) ratios | | (***S/D***)-***error-ratio*** $\left\lvert \frac{^{Z}Đ_\delta}{^{Z}Đ_A} \right\rvert$ | $^{S}\boldsymbol{R_i}$-value (cf. either Eq. 10 or 10a) | ***Error ratio*** as $\left\lvert \frac{^{Z}Đ_\delta}{^{S}Đ_i} \right\rvert$ |
|---|---|---|---|---|---|---|---|---|
| | | Estimated (***S/W***)-ratio "$y_A$" (Error: $Đ_A \times 10^2$) | (***S/D***)-estimate "$z_A$" (Error: $^{Z}Đ_A \times 10^2$) | $^{S/W}$**δ**-ratio "$y_\delta$" (Error: $Đ_\delta \times 10^2$) | $^{S/D}$**δ**-ratio "$z_\delta$" (Error: $^{Z}Đ_\delta \times 10^2$) | | $^{S}r_i \times 10^5$ (Error: $^{S}Đ_i \times 10^2$) | |
| 1 | 0 | 1.1103336 (0) | 0.8996700 (0) | 0.1103336 (0) | −0.10033 | 0 | 14.013260 (0) | 0 |
| | 1 | 1.1104335 (0.009) | 0.89975097 (0.009) | 0.1104335 (0.0906) | −0.1002490 (−0.08070) | 8.9671 | 14.014521 (0.009) | 8.9671 |
| | 2 | 1.1100117 (−0.029) | 0.8994092 (−0.029) | 0.1100117 (−0.292) | −0.1005908 (0.26) | 8.9671 | 14.009197 (−0.029) | 6.9671 |
| | 3 | 1.1102670 (−0.006) | 0.89961603 (−0.006) | 0.1102670 (−0.0604) | -0.1003840 (0.0538) | 8.9671 | 14.012419 (−0.006) | 8.9671 |
| | 4 | 1.1107112 (0.034) | 0.89997595 (0.034) | 0.1107112 (0.3422) | −0.1000241 (−0.3049) | 8.9671 | 14.018026 (0.034) | 8.9671 |
| | 5 | 1.1101781 (−0.014) | 0.89954405 (−0.014) | 0.1101781 (−0.1409) | −0.1004560 (0.1255) | 8.9671 | 14.011298 (−0.014) | 8.9671 |
| 2 | 0 | 1.1103336 (0) | 0.8996700 (0) | 0.1103336 (0) | −0.10033 (0) | 0 | 14.013260 (0) | 0 |
| | 1 | 1.1104336 (0.009) | 0.89975104 (0.00901) | 0.1104336 (0.0906) | −0.1002490 (−0.08077) | 8.9671 | 14.014522 (0.00901) | 8.9671 |
| | 2 | 1.1100225 (−.02802) | 0.89941791 (−0.02802) | 0.1100225 (−0.282) | −0.1005821 (0.2513) | 8.9671 | 14.009333 (−0.02802) | 8.9671 |
| | 3 | 1.1102558 (−0.0070) | 0.89960698 (−0.0070) | 0.1102558 (−0.0705) | −0.1003932 (0.0628) | 8.9671 | 14.012278 (−0.0070) | 8.9671 |
| | 4 | 1.1107338 (0.03604) | 0.89999424 (0.03604) | 0.1107338 (0.3627) | −0.1000058 (−0.3232) | 8.9671 | 14.018310 (0.03604) | 8.9671 |
| | 5 | 1.1101113 (−.02002) | 0.89948989 (−0.02002) | 0.1101113 (−0.2015) | −0.1005101 (0.17952) | 8.9671 | 14.010455 (−0.02002) | 8.9671 |